%% file: main.tex
\documentclass[letterpaper, 10pt, conference]{ieeeconf}

\IEEEoverridecommandlockouts

\usepackage{amsmath}  %
\usepackage{amsfonts,amssymb}
\usepackage{xspace}    %

\usepackage{graphicx}
\graphicspath{{figs/}}

\usepackage[caption=false,font=footnotesize]{subfig}

\usepackage{pgfplots}
\usepackage{grffile}
\pgfplotsset{compat=newest}  %
\usetikzlibrary{plotmarks}
\usetikzlibrary{arrows.meta}
\usepgfplotslibrary{patchplots}

\newlength{\figheight}
\newlength{\figwidth}

\usepackage{tabularx}

\usepackage[ruled]{algorithm}
\usepackage{algpseudocode}

\makeatletter
\newcommand{\multiline}[1]{%
  \begin{tabularx}{\dimexpr\linewidth-\ALG@thistlm}[t]{@{}X@{}}
    #1
  \end{tabularx}
}
\makeatother

\usepackage[per-mode=symbol]{siunitx}

\usepackage{cite}  %
\usepackage{array}  %
\usepackage{booktabs}           %

\newtheorem{assumption}{Assumption}

\newtheorem{remark}{Remark}

\newcounter{problem}
{\par\endtrivlist\unskip}

\usepackage[reviewmark,extramath,probability]{truonglatexdefs}
\def\linGP{linGP\xspace}

\newcommand{\GPModel}[1][f]{\ensuremath{\GGG_{#1}}}

\newcommand{\GPmean}[1]{\ensuremath{\mu_{#1}}}

\newcommand{\GPvar}[1]{\ensuremath{\sigma_{#1}^{2}}}

\newcommand{\linGPModel}[2]{\ensuremath{\tilde{\GGG}_{#1 | #2}}}

\newcommand{\linGPmean}[2]{\ensuremath{\tilde{\mu}_{#1 | #2}}}

\newcommand{\linGPvar}[2]{\ensuremath{\tilde{\sigma}_{#1 | #2}^{2}}}

\renewcommand{\IdentityMatrix}{\mathrm{I}}  %

\title{Linearized Gaussian Processes for Fast Data-driven Model Predictive Control}

\author{Truong X. Nghiem\\
  School of Informatics, Computing, and Cyber Systems\\
  Northern Arizona University\\
  truong.nghiem@nau.edu %
}

\begin{document}

\maketitle
\thispagestyle{empty}
\pagestyle{empty}

\begin{abstract}
  Data-driven Model Predictive Control (MPC), where the system model is learned from data with machine learning, has recently gained increasing interests in the control community. %
  Gaussian Processes (GP), as a type of statistical models, are particularly attractive due to their modeling flexibility and %
  their ability to provide probabilistic estimates of prediction uncertainty.
  GP-based MPC has been developed and applied, however the optimization problem is typically non-convex and highly demanding, and scales poorly with model size.
  This causes unsatisfactory solving performance, even with state-of-the-art solvers, and makes the approach less suitable for real-time control.
  We develop a method based on a new concept, called \emph{linearized Gaussian Process}, and Sequential Convex Programming, that can significantly improve the solving performance of GP-based MPC.
  Our method is not only faster but also much more scalable and predictable than other commonly used methods, as it is much less influenced by the model size.
  The efficiency and advantages of the algorithm are demonstrated clearly in a numerical example.
\end{abstract}

\section{Introduction}
\label{sec:introduction}
\input{intro}

\section{Linearized Gaussian Processes (linGP)}
\label{sec:lingp}
\input{lingp}
\section{GP-based Model Predictive Control}
\label{sec:gp-mpc}
\input{gpmpc}

\section{linGP-based Model Predictive Control}
\label{sec:lingp-mpc}
\input{lingpmpc}

\section{Numerical Example}  %
\label{sec:simulation}
\input{simulation}

\section{Conclusion}
\label{sec:conclusions}
\input{conclusion}

\bibliographystyle{IEEEtran}
\bibliography{IEEEabrv,references}

\end{document}

%% file: intro.tex
Model predictive control (MPC) is an advanced control approach that utilizes a mathematical model of the controlled process to predict its future responses over a finite time horizon, then, based on these predictions, minimizes a cost function to obtain optimized control inputs subject to input and state constraints.
This procedure is repeated at every time step, resulting in an optimal control strategy.
The MPC approach is attractive due to its ability to effectively and intuitively handle complex dynamics and system constraints \cite{maciejowski_predictive_2002}.
Stochastic MPC (SMPC) %
can efficiently handle stochastic uncertainties with underlying probability distributions in optimal control problems \cite{schwarm1999Chanceconstrainedmodelpredictive}.
For these reasons, MPC and its variants have been applied widely in many practical control problems \cite{mayne2014Modelpredictivecontrol}.

A major caveat of MPC is that it requires a reasonably accurate mathematical model of the system because its performance is highly dependent on the accuracy of the model-based predictions.
Traditionally, these models are developed using first principles based on physics.
However, for complex systems where knowledge of the system dynamics is often incomplete or hard to obtain fully, the required effort for such model development and engineering would pose significant difficulty and limit the use of physics-based models for MPC \cite{JainICCPS2018}.
To overcome this challenge, data-driven modeling for MPC has recently been studied.
This approach uses machine learning techniques to learn a system model from data, possibly with certain prior knowledge about the system, and uses that model in place of the traditional mathematical model in MPC.
Such \emph{data-driven MPC} approach has become increasingly appealing, especially for complex and large-scale systems, as a result of recent advancements in machine learning, optimization, and computation.

Gaussian Processes (GPs) -- a type of statistical machine learning models -- have been used for modeling dynamical systems and for MPC \cite{Kocijan2016}.
GPs are highly flexible and able to capture complex behaviors with fewer parameters than other machine learning techniques, hence they generally work well with small data sets \cite{Rasmussen2006}.
More importantly, a GP provides an \emph{estimate of uncertainty or doubt} in the predictions through the predictive variance, which can be used to assess or guarantee the performance of a learning-based system.
Because of these advantages, GPs have been employed successfully in MPC frameworks and applications %
\cite{Kocijan2016,nghiemetal16gp,JainICCPS2018,cao2017Gaussianprocessmodel}.
In this paper, we will consider the GP-based MPC approach for data-driven predictive control, which we will call GP-MPC henceforth.

Computational complexity is a major challenge of GPs, thus of GP-MPC.
GP computations generally scale cubically with the size of the training data \cite{Rasmussen2006}.
When the training data size increases, GP-MPC quickly becomes very expensive to solve.
This is particularly true when the GP variance, which represents model uncertainty, is %
used in the cost function or the constraints, leading to a non-convex and highly demanding optimization problem.
As a result, a general-purpose nonlinear program (NLP) solver is usually used to solve GP-MPC, for instance by Interior-Point methods (IPM), Sequential Quadratic Programming (SQP) methods, and active-set methods \cite{Kocijan2016}. %
However, the performance of these %
methods in GP-MPC problems is usually unsatisfactory.
For instance, it was reported in \cite{nghiemetal16gp} that a moderate-sized GP-MPC problem took up to a minute to solve with Ipopt.
Section~\ref{sec:simulation} reports the solving time of three general NLP methods for a numerical GP-MPC example, illustrating the computational performance issues of these methods.

Our goal is to make GP-MPC fast, scalable, and more predictable compared to other NLP methods. %
In particular, we aim to make solving GP-MPC less influenced by the training data size.
We propose a new concept called \emph{linearized Gaussian Process} (linGP), which is a (valid) Gaussian Process of the linearization of the latent function about a given input.
This is not the same as linearizing the GP mean and variance because a \linGP is still a GP and therefore it has mean, variance, and moments of any order.
We then develop a Sequential Convex Programming (SCP) algorithm based on \linGP to solve a general class of GP-MPC problems very efficiently and, most importantly, much more independently of the training data size compared to other methods.
The proposed SCP algorithm is inspired by the successive convexification approach in \cite{mao2016Successiveconvexificationnonconvex} to solve non-convex optimal control problems and similar in concept to successive linearization methods to solve nonlinear MPC (such as in \cite{lawrynczukComputationallyEfficientModel2014}).
However, our approach is different from other attempts to accelerate GP-MPC and machine-learning-based MPC in general, for example \cite{cao2017Gaussianprocessmodel,lawrynczukComputationallyEfficientModel2014}, in that we approximate not the output equation of the machine learning model but the underlying latent process of the model.   %

The concept of \linGP is introduced in Section~\ref{sec:lingp}, followed by the considered class of GP-MPC problems in Section~\ref{sec:gp-mpc}.
We present the \linGP-based SCP algorithm in Section~\ref{sec:lingp-mpc}.
Finally, the efficacy of our method is demonstrated in a numerical example in Section~\ref{sec:simulation}.

%% file: lingp.tex
This section introduces the concept of a \emph{linearized Gaussian Process} (\emph{\linGP}) and develops the mathematical derivation of a \linGP from a given Gaussian Process (GP).
We will begin with a brief introduction to GPs and the GP-related notations used in this paper.

\subsection{Gaussian Processes: Basics and Notations}
\label{sec:gp-basics}

Given $N$ noisy observations \(y^{(i)}\) of an underlying function \(f: \RR^n \mapsto \RR\) through a Gaussian noise model: \(y^{(i)} = f\left(x^{(i)}\right) + \epsilon^{(i)}\), with inputs \(x^{(i)} \in \RR^n\), \(\epsilon^{(i)} \sim \GaussianDist{0}{\sigma_n^2}\), and $i = 1,\dots,N$ .
A GP of $f$, considered as a distribution over functions \cite{Rasmussen2006}, is essentially a probability distribution on the observations of all possible realizations of $f$.
The \emph{prior} $\PP(f)$ represents the initial belief about this distribution.
Let $\DDD = (X, Y)$ be the finite set of observation data of $f$, where \(X = [x^{(1)}, \dots, x^{(N)}] \in \RR^{n \times N}\) collects the regression vectors and \(Y = [y^{(1)}, \dots, y^{(N)}] \in \RR^{N}\) contains the corresponding observed outputs.
By conditioning the GP on $\DDD$, the \emph{posterior} $\PP(f | \DDD)$ is the updated distribution after seeing the observation data, which allows probabilistic inference at a new input.

The GP assigns a joint Gaussian distribution to any finite subset of random variables $\{f^{(1)}, \dots, f^{(M)}\}$ corresponding to inputs $\{x^{(1)}, \dots, x^{(M)}\}$.
It is fully specified by its mean function \(m(x)\) and covariance function \(k(x,x')\), parameterized by the \emph{hyperparameters} \(\theta\),
\begin{align*}
m(x; \theta) &= \EE [f(x)] \\
k(x,x'; \theta) &= \EE [(f(x)\!-\!\mu(x)) (f(x') \!-\! \mu(x'))] \text.\nonumber %
\end{align*}
The mean function is %
often set to zero without loss of generality.

The output \(f_\star\) of the GP corresponding to an input \(x_\star\) is a random variable \(f_{\star} \sim \GaussianDist{\mu_\star}{\sigma_\star^2}\), where
\begin{subequations}
\label{eq:gp-regression}
\begin{align}
\mu_\star &= m(x_\star) + K_\star K^{-1} (Y - \mu(X)) \label{eq:gp-regression:mean} \\
\sigma_\star^2 &= K_{\star \star} - K_\star K^{-1} K_\star^T \text,\label{eq:gp-regression:var}
\end{align}
\end{subequations}
in which \(K_\star = [k(x_\star, x^{(1)}), \dots, k(x_\star, x^{(N)})] \in \RR^{N}\), \(K_{\star \star} = k(x_\star, x_\star) \in \RR\), and $K \in \RR^{N \times N}$ is the covariance matrix with elements \(K_{ij} = k(x^{(i)}, x^{(j)})\).
Note that the mean and variance are nonlinear in $x_{\star}$ and their computations scale cubically with the size of $X$ and $Y$.

In theory, %
the hyperparameters $\theta$ are also random variables, %
whose posterior distributions are obtained by conditioning them on $\DDD$ using the Bayes' theorem.
In practice, however, $\theta$ are often obtained by maximizing the likelihood: \(\argmax_\theta \Pr(Y \vert X, \theta)\).
For an in-depth treatment of GPs, interested readers are referred to \cite{Rasmussen2006}.

\subsection{Linearized Gaussian Processes}
\label{sec:lingp-def}

Consider a GP of a nonlinear function $f(x): \RR^{n} \to \RR$, with covariance function \(k(x, x') : \RR^{n} \times \RR^{n} \to \RR\).
We will denote the GP of $f$ by \GPModel[f], whose predictive mean and variance at an input $x \in \RR^{n}$ are given by \eqref{eq:gp-regression} and are denoted by $\GPmean{f}(x)$ and $\GPvar{f}(x)$, respectively.
For notational simplicity we will take the mean function $m(x)$ to be zero.
If the GP has a non-zero mean function, it is equivalent to the sum of %
$m(x)$ and a zero-mean GP, hence our %
results can %
be extended to this case.

The GP inference at any input $x$, given by \eqref{eq:gp-regression}, is non-convex due to the covariance function $k(\cdot, \cdot)$.
This makes any optimization %
involving GP regressions %
non-convex and generally difficult to solve.
However, in many of these problems, non-convexity only comes from the GP regressions involved.
One typical approach to overcome this issue is to approximate the mean and variance %
by linear or quadratic equations using Taylor's series expansions, which often results in an approximated linear or quadratic program.
For instance, this approach is employed by SQP solvers.
However, as shown in Section~\ref{sec:introduction}, %
this approach often does not work well.
In this paper, we take a different approach, described below.

Given an input vector $x$, we assume that $f$ is differentiable at $x$.
We also assume that the covariance function $k(\cdot, \cdot)$ is differentiable, which is typically the case \cite{solak2003DerivativeObservationsGaussian}.
Recall that the GP of $f$ is essentially a (posterior) distribition over the space of realizations of %
$f$. %
We can then linearize %
$f$ around $x$ as %
\begin{equation*}
  f(x + \Delta_{x}) \approx \tilde{f}_{x}(\Delta_{x}) = f(x) + \Delta_{x}^{T} \nabla_{x} f(x) , \quad \Delta_{x} \in \RR^{n }\text.
\end{equation*}
Define $g \equiv \nabla_{x} f$ as the gradient of $f$.
The above linearization can be rewritten as, with $\hat{x} = [1, \Delta_{x}^{T}]^{T}$,
\begin{equation}
  \label{eq:gp-linearized}
  \tilde{f}_{x}(\Delta_{x}) =
  \begin{bmatrix}
    1 \\
    \Delta_{x}
  \end{bmatrix}^{T}
  \begin{bmatrix}
    f(x) \\
    g(x)
  \end{bmatrix} =
  \hat{x}^{T} \hat{f}(x) \text.
\end{equation}
It is important to note here that
$\hat{f}(x) = [f(x), g^{T}(x)]^{T}$
is a random vector of length $(n+1)$ which, through the inner product with the vector $\hat{x}$, results in the random variable $\tilde{f}_{x}(\Delta_{x})$ of the linearized process.
In other words, \emph{we approximate the original GP of ${f}$ around ${x}$ by a process of linearized function ${\tilde{f}_{x}}$ of ${f}$, which we will call a \emph{linearized Gaussian Process} or \emph{linGP}}.
The linGP of the GP \GPModel[f] at input $x$ is denoted by $\linGPModel{f}{x}$.

As differentiation is a linear operator, $\hat{f}$ is a (multivariate) Gaussian Process derived from the original GP.
Therefore, $\hat{f}(x)$ is a Gaussian random vector that defines a posterior distribution, derived from the posterior $\PP(f | X,Y)$, over the values of $f(x)$ and its gradiant at $x$.
Consequently, the linGP \linGPModel{f}{x} %
has a distribution derived from $\PP(f | X,Y)$ over the linear approximations of the function $f$ near $x$ .
To characterize this linGP, %
we need to characterize the GP of $\hat{f}$.

Let $K^{(1,0)} = (\nabla_{x} k)$ -- the gradient of $k$ with respect to the first argument -- and $K^{(0,1)} = (D_{x'} k)$ -- the Jacobian of $k$ with respect to the second argument.
Let $K^{(1,1)}(x,x')$ be the $n \times n$ matrix such that $K^{(1,1)}_{i,j}(x,x') = \frac{\partial^{2}}{\partial x_{i} \partial x'_{j}} k(x,x')$.
Following \cite{solak2003DerivativeObservationsGaussian}, we can write the joint distribution of the observation data, the process $f$, and the gradient process $g$ as
\begin{equation}
  \label{eq:full-joint-distribution}
  \left[\begin{matrix}
    Y \\ f \\ g
  \end{matrix}\right] \!\sim\!
  \GaussianDist{0}{%
    \left[\begin{smallmatrix}
        K(X,X) + \sigma_{n}^{2} \IdentityMatrix & k(X,x) & K^{(0,1)}(X,x) \\
        k(x,X) & k(x,x) & K^{(0,1)}(x,x) \\
        K^{(1,0)}(x,X) & K^{(1,0)}(x,x) & K^{(1,1)}(x,x)
      \end{smallmatrix}\right]} \text.
\end{equation}
Here, we use the convention that when a function is applied with $X$, it is broadcast along the corresponding dimensions.
For example, $K(X,X)$ is a $N \times N$ matrix with $K_{i,j}(X,X) = k(x^{(i)}, x^{(j)})$; similarly, the $i^{\mathrm{th}}$ row of $K^{(0,1)}(X,x)$ is $K^{(0,1)}(x^{(i)}, x)$.
The posterior distribition of $\hat{f}$ conditioned on the observation data $(X,Y)$ is then
\begin{gather}
  \label{eq:derivative-distribution}
  \hat{f} | X,Y,x \sim \GaussianDist{\hat{m}_x}{\hat{V}_x} \\
  \hat{m}_x =
  \begin{bmatrix}
    k(x,X) \\ K^{(1,0)}(x,X)
  \end{bmatrix}
  \left( K(X,X) + \sigma_{n}^{2} \IdentityMatrix \right)^{-1} Y \nonumber \\
  \begin{aligned}
    \hat{V}_x = &
    \left[\begin{smallmatrix}
        k(x,x) & K^{(0,1)}(x,x) \\
        K^{(1,0)}(x,x) & K^{(1,1)}(x,x)
      \end{smallmatrix}\right] - \\
    &
    \left[\begin{smallmatrix}
        k(x,X) \\ K^{(1,0)}(x,X)
      \end{smallmatrix}\right]
    \scriptstyle\left( K(X,X) + \sigma_{n}^{2} \IdentityMatrix \right)^{-1}
    \left[\begin{smallmatrix}
        k(X,x) & K^{(0,1)}(X,x)
      \end{smallmatrix}\right]
  \end{aligned} \nonumber
\end{gather}
where $\IdentityMatrix$ denotes an identity matrix of appropriate dimensions.
The linearized process in \eqref{eq:gp-linearized} can then be written as follows.
\begin{equation}
  \label{eq:gp-linearized-final}
  \tilde{f}_{x}(\Delta_{x}) \sim \NNN \left(
    \hat{m}_x^{T} \hat{x},
    \hat{x}^{T} \hat{V}_x \hat{x}
  \right) \text.
\end{equation}
The mean of the original GP at $(x + \Delta_{x})$ is approximated by a linear function of $\hat{x}$, while its variance is approximated by a quadratic function of $\hat{x}$.
Because $\hat{f}$ is a valid GP, $\hat{V}_x$ is positive semi-definite (PSD) and, therefore, it is guaranteed that the %
variance of $\tilde{f}_{x}(\Delta_{x})$ is non-negative.
We emphasize again that \emph{the linGP is in fact a Gaussian Process of ${\tilde{f}_{x}}$, which approximates the original GP around ${x}$.
A key advantage of the linGP is that its mean and variance are respectively linear and convex quadratic functions.}
We will use $\linGPmean{f}{x}(\Delta_{x})$ and $\linGPvar{f}{x}(\Delta_{x})$ as the notations for the mean and variance of the linGP \linGPModel{f}{x} at input $(x+\Delta_{x})$ in a neighborhood of $x$.
They are calculated as
\begin{equation}
  \label{eq:lingp-mean-var}
  \linGPmean{f}{x}(\Delta_{x}) = \hat{m}_x^{T} \hat{x}, \quad
  \linGPvar{f}{x}(\Delta_{x}) = \hat{x}^{T} \hat{V}_x \hat{x} \text,
\end{equation}
in which $\hat{m}_{x}$ and $\hat{V}_{x}$ are given in \eqref{eq:derivative-distribution}, and $\hat{x}$ is defined in \eqref{eq:gp-linearized}.

\begin{remark}
  The above derivations can easily be extended to the case when a subset of elements in $x$ are fixed.
  In that case, the elements of $\hat{x}$, $K$, $K^{(0,1)}$, $K^{(1,0)}$, and $K^{(1,1)}$ corresponding to the fixed elements of $x$ are removed.
\end{remark}

\begin{remark}
  The mean and variance of the predictive output of the GP can be approximated directly based on Taylor series, as commonly employed by SQP-based NLP methods.
  A key distinction between such an approximation approach and \linGP is that while the former linearizes the mean and variance of the probabilistic predictive output (\ie linearization happens \emph{after} prediction), \linGP linearizes the latent function directly and derives the probabilistic distribution of such linearized functions (\ie linearization happens \emph{before} prediction).
  The mean $\linGPmean{f}{x}$ obtained above is exactly the same as if the exact mean equation \eqref{eq:gp-regression:mean} is linearized directly.
  However, the variance $\linGPvar{f}{x}$ is not the same as if the exact variance equation \eqref{eq:gp-regression:var} is approximated to the second order by a Taylor series.
  Indeed, while $\linGPvar{f}{x}$ is a true variance and guaranteed to be non-negative, %
  it is not necessarily the case with a second-order Taylor series approximation.
  As we will explain later in Section~\ref{sec:complexity-lingp-mpc}, this key difference is an important advantage of the \linGP approach, that could lead to a substantial performance improvement of our algorithm compared to other methods.  
\end{remark}

%% file: gpmpc.tex
In this section, we will consider a general %
MPC formulation using %
GP models.

\subsection{Gaussian Processes for Dynamical Systems}
\label{sec:dynamical-gp}

GPs can be used for modeling nonlinear dynamical systems \cite{Kocijan2016}.
This can be achieved by feeding \emph{autoregressive}, or time-delayed, input and output signals back to the model as regressors.
Specifically, it is common %
to model a discrete-time dynamical system by an autoregressive nonlinear function
\begin{equation}
  \label{eq:dynamical-gp:autoregressive-function}
y_{k} = f(y_{k-l}, \dots, y_{k-1}, u_{k-m}, \dots, u_k) \text. %
\end{equation}
Here, \(k\) denotes the time step, \(u \in \RR^{n_{u}}\) the control input, %
\(y \in \RR\) the output, and \(l\) and \(m\) %
are respectively the lags for autoregressive outputs and control inputs. %
The regressor vector includes not only the current control input \(u_k\) %
but also the past (\ie autoregressive) values of $y$ and $u$. %
The vector of all autoregressive inputs can be thought of as the current state of the model.
The nonlinear function $f$ of the dynamical system can be learned by a GP \GPModel[f], trained from the system's data in the same way as any other GPs.
MPC formulations that use GP models of the involved dynamical systems will be called GP-MPCs in this paper.
We assume that a single GP is used in a GP-MPC, however our results readily extend to the case when multiple GPs are involved.

Due to the predictive nature of a GP-MPC, it is necessary to simulate the GP \GPModel[f] over $H$ future steps, where $H$ is the MPC horizon, and predict its multistep-ahead behavior.
Because the output of \GPModel[f] is a distribution rather than a point estimate, the autoregressive outputs fed to the model beyond the first step are random variables, resulting in more and more complex output distributions as we go further.
Therefore, a multistep simulation of \GPModel[f] involves the propagation of uncertainty through the model.
There exist several methods for approximating uncertainty propagation in GPs \cite{Kocijan2016}.

It was shown in \cite{nghiemetal16gp} that the \emph{zero-variance method}, which replaces the autoregressive outputs with their corresponding expected values and therefore does not propagate uncertainty, could achieve sufficient prediction accuracy. %
Its computational simplicity is attractive, especially in GP-MPC where the GP must be simulated for many time steps.
In this paper, we will assume the zero-variance method for predicting future GP outputs in MPC.
Let us define the state vector $x$ such that, at time step $k$,
\begin{multline}
  \label{eq:dynamical-gp:state-vector}
  x_{k}^{T} = [\overbar{y}_{k-l}, \overbar{y}_{k-l+1},\dots, \overbar{y}_{k-1}, \qquad \\
  u_{k-m}^{T}, u_{k-m+1}^{T}, \dots, u_{k-1}^{T}] %
\end{multline}
where $\overbar{y}_{i}$ is the expected value of $y_{i}$.
Note that all $\overbar{y}_{k}$ for $k < t$, where $t$ is the current time, can be observed. %

The output of \GPModel[f] at time $k \geq t$ then reads
\begin{subequations}
\begin{gather}
  \label{eq:dynamical-gp:gp-output}
  y_{k} \sim \GaussianDist{\overbar{y}_{k}}{\sigma_{y,k}^{2}} \\
  \overbar{y}_{k} = \GPmean{f}(x_{k}, u_{k}), \quad  %
  \sigma_{y,k}^{2} = \GPvar{f}(x_{k}, u_{k}) \text. %
\end{gather}
\end{subequations}

\subsection{Model Predictive Control with Gaussian Process Models}
\label{sec:mpc-formulation}

In this paper, we consider a general GP-MPC formulation described below.
Let $t$ be the current time step and $H > 0$ the MPC horizon length.
Define the collections $\overbar{\YYY}_{t} = \{\overbar{y}_{k} \,|\, k \in \III_{t}\}$, $\SSS_{y,t} = \{\sigma_{y,k} \,|\, k \in \III_{t}\}$, and $\UUU_{t} = \{u_{k} \,|\, k \in \III_{t}\}$, %
where $\III_{t} = \{t,\dots,t+H-1\}$ contains the indices of all time steps in the current MPC horizon.
We also introduce additional variables $z \in \RR^{n_{z}}$ that are used in the MPC formulation but not in the GP model, for example to model non-GP dynamics in the system.
The collection of these variables in the MPC horizon is denoted by $\ZZZ_{t} = \{z_{k} \,|\, k \in \III_{t}\}$.

The GP-MPC reads
\begin{subequations}
\label{eq:gpmpc}
\begin{align}
  & \text{\bfseries (GP-MPC)} \nonumber\\  
  & \underset{\UUU_{t}, \ZZZ_{t}}{\text{minimize}} \quad J\left(\overbar{\YYY}_{t}, \SSS_{y,t}, \UUU_{t}, \ZZZ_{t}\right) \label{eq:gpmpc:obj}\\
  & \text{subject to} \nonumber \\
  & \qquad \overbar{y}_{k} = \GPmean{f}(x_{k}, u_{k}) \label{eq:gpmpc:mean}\\  %
  & \qquad \sigma_{y,k}^{2} = \GPvar{f}(x_{k}, u_{k}) \label{eq:gpmpc:var}\\  %
  & \qquad z_{k} \in Z, \quad u_{k} \in U \label{eq:gpmpc:varconstraints}\\
  & \qquad g_{i} \left(\overbar{\YYY}_{t}, \SSS_{y,t}, \UUU_{t}, \ZZZ_{t}\right) \leq 0, \, i = 1,\dots,n_{\mathrm{ieq}} \label{eq:gpmpc:ineqs}\\
  & \qquad h_{i} \left(\overbar{\YYY}_{t}, \SSS_{y,t}, \UUU_{t}, \ZZZ_{t}\right) = 0, \, i = 1,\dots,n_{\mathrm{eq}} \label{eq:gpmpc:eqs}
\end{align}
\end{subequations}
in which $k$ ranges from $t$ to $t+H-1$ and
\begin{itemize}
\item $J(\cdot)$ is the objective function and typically has the form
  \(J = \textstyle\sum_{k=t}^{t+H-1} c_{k}\left( \overbar{y}_{k}, \sigma_{y,k}, u_{k}, z_{k}\right)\)
  where $c_{k}(\cdot)$ are the stage cost functions;
\item Constraints~\eqref{eq:gpmpc:mean} and \eqref{eq:gpmpc:var} represent the GP dynamics;
\item Sets $Z$ and $U$ in \eqref{eq:gpmpc:varconstraints} are convex constraint sets of the variables $z$ and $u$, respectively;
\item There are $n_{\mathrm{ieq}}$ inequality constraints of the form \eqref{eq:gpmpc:ineqs};
\item There are $n_{\mathrm{eq}}$ equality constraints of the form \eqref{eq:gpmpc:eqs};
\end{itemize}

We make the following assumptions about the GP-MPC formulation~\eqref{eq:gpmpc}.
\begin{assumption}
  \label{assumption:convexity}
  Suppose that, for every $k$, $\overbar{y}_{k}$ is affine and $\sigma_{y,k}^{2}$ is convex quadratic in the control variables $\UUU_{t}$.
  Then $J$ is convex, each $g_{i}$ is convex, and each $h_{i}$ is affine in the optimization variables $\UUU_{t}$ and $\ZZZ_{t}$.
\end{assumption}
This assumption holds in many applications of GP-MPC because the non-convexity of \eqref{eq:gpmpc} usually comes solely from the GP dynamics \cite{Kocijan2016,nghiemetal16gp,JainICCPS2018}.
It also usually holds in stochastic GP-MPC %
with chance constraints approximated by convex deterministic constraints (see, \eg \cite{nghiemetal16gp,JainICCPS2018}).

Because of the non-convexity of the general GP-MPC %
\eqref{eq:gpmpc}, a non-convex NLP %
solver %
is usually employed.
However, as discussed in Section~\ref{sec:introduction} and illustrated in Section~\ref{sec:simulation}, the performance of these solvers on GP-MPC problems are usually unsatisfactory, especially in time-critical control applications.

%% file: lingpmpc.tex
In this section, an algorithm based on \linGP is developed to solve the GP-MPC problem \eqref{eq:gpmpc} with significantly improved performance compared to general %
NLP solvers.
More importantly, %
its complexity is much less influenced by the size of the GP model (\ie the size of the training data set).

\subsection{Local GP-MPC problem with \linGP}
\label{sec:local-mpc-problem}

We first rewrite the GP-MPC problem in an equivalent form.
Suppose that nominal control inputs $u^{\star}_{k}$, for $k \in \III_{t} = \{t,\dots,t+H-1\}$, are given.
One can then simulate the GP model \GPModel[f] to obtain the nominal outputs $\overbar{y}^{\star}_{k}$.
As the vectors $x_{k}$ defined in \eqref{eq:dynamical-gp:state-vector} are constructed from the control inputs and GP outputs, we also obtain their nominal values $x^{\star}_{k}$.
We note that the exact control inputs and process outputs before the current time $t$ are available (calculated or measured).
With these nominal values, define $u_{k} = u^{\star}_{k} + \Delta u_{k}$ and $x_{k} = x^{\star}_{k} + \Delta x_{k}$, for all $k \in \III_{t}$.
All $\Delta u_{k}$ are collected in $\Delta\UUU_{t} = \{\Delta u_{k} \,|\, k \in \III_{t}\}$.
With a slight abuse of notations, we will write $\UUU_{t} = \UUU^{\star}_{t} + \Delta \UUU_{t}$, where $\UUU^{\star}_{t}$ collects all nominal control inputs.
The GP-MPC formulation \eqref{eq:gpmpc} is then equivalent to
\begin{subequations}
\label{eq:gpmpc-delta}
\begin{align}
  & \underset{\Delta \UUU_{t}, \ZZZ_{t}}{\text{minimize}} \quad J\left(\overbar{\YYY}_{t}, \SSS_{y,t}, \UUU^{\star}_{t} + \Delta\UUU_{t}, \ZZZ_{t}\right) \label{eq:gpmpc-delta:obj}\\
  & \text{subject to} \nonumber \\
  & \quad \overbar{y}_{k} = \GPmean{f}(x^{\star}_{k} + \Delta x_{k}, u^{\star}_{k}+\Delta u_{k}) \label{eq:gpmpc-delta:mean}\\  %
  & \quad \sigma_{y,k}^{2} = \GPvar{f}(x^{\star}_{k} + \Delta x_{k}, u^{\star}_{k}+\Delta u_{k}) \label{eq:gpmpc-delta:var}\\  %
  & \quad z_{k} \in Z, \quad u^{\star}_{k}+\Delta u_{k} \in U \label{eq:gpmpc-delta:varconstraints}\\
  & \quad g_{i} \left(\overbar{\YYY}_{t}, \SSS_{y,t}, \UUU^{\star}_{t} + \Delta\UUU_{t}, \ZZZ_{t}\right) \leq 0, \, i = 1,\dots,n_{\mathrm{ieq}} \label{eq:gpmpc-delta:ineqs}\\
  & \quad h_{i} \left(\overbar{\YYY}_{t}, \SSS_{y,t}, \UUU^{\star}_{t} + \Delta\UUU_{t}, \ZZZ_{t}\right) = 0, \, i = 1,\dots,n_{\mathrm{eq}} \label{eq:gpmpc-delta:eqs}
\end{align}
\end{subequations}

With this equivalent formulation, the GP-MPC problem can be approximated locally around the nominal values by replacing the GP model \GPModel[f] with its \linGP \linGPModel{f}{(x^{\star}_{k}, u^{\star}_{k})} at each $k$.
Specifically, $\overbar{y}_{k}$ and $\sigma^{2}_{y,k}$ are replaced by the mean $\tilde{y}_{k}$ and variance $\tilde{\sigma}^{2}_{y,k}$ of the \linGP at nominal inputs $x^{\star}_{k}$ and $u^{\star}_{k}$, as
\(\tilde{y}_{k} = \linGPmean{f}{x^{\star}_{k}, u^{\star}_{k}}(\Delta x_{k}, \Delta u_{k})\)
and
\(\tilde{\sigma}_{y,k}^{2} = \linGPvar{f}{x^{\star}_{k}, u^{\star}_{k}}(\Delta x_{k}, \Delta u_{k})\)
respectively.
It follows from the derivation of the linGP, %
particularly Eqs.~\eqref{eq:lingp-mean-var}, that $\tilde{y}_{k}$ is linear and $\tilde{\sigma}^{2}_{y,k}$ is convex quadratic in $\Delta x_{k}$ and $\Delta u_{k}$:
\(  \tilde{y}_{k} = \hat{m}^{T}_{x^{\star}_{k}, u^{\star}_{k}} \xi_{k} \) %
and
\(  \tilde{\sigma}_{y,k}^{2} = \xi_{k}^{T} \hat{V}_{x^{\star}_{k}, u^{\star}_{k}} \xi_{k}\),
where
$\xi_{k}^{T} = \left[ 1, \Delta x_{k}^{T}, \Delta u_{k}^{T}\right]$.
The vector $\hat{m}$ and the %
matrix $\hat{V}$ are defined in \eqref{eq:derivative-distribution}.
Let $\Delta \tilde{y}_{k}=\tilde{y}_{k} - \overbar{y}^{\star}_{k}$.
Then $\Delta x_{k}$ is derived from $\Delta \tilde{y}_{i}$ and $\Delta u_{i}$ for $i < k$, see \eqref{eq:dynamical-gp:state-vector}, as
\begin{multline}
  \label{eq:diff-state-vector}
  \Delta x_{k}^{T} = [\Delta \tilde{y}_{k-l}, \Delta\tilde{y}_{k-l+1},\dots, \Delta\tilde{y}_{k-1}, \qquad \\
  \Delta u_{k-m}^{T}, \Delta u_{k-m+1}^{T}, \dots, \Delta u_{k-1}^{T}]\text. %
\end{multline}

As the approximate problem, in particular the \linGP model, is only %
accurate in a neighborhood of the nominal GP inputs $x^{\star}_{k}$ and $u^{\star}_{k}$, we introduce a \emph{trust region} of size $\rho \geq 0$.
Specifically, we impose the bounds, for all $k \in \III_{t}$
\begin{equation}
  \label{eq:lingpmpc-trustregion}
  \norm{\Delta u_{k}} \leq \rho, \quad \norm{\Delta \tilde{y}_{k}} \leq \rho
\end{equation}
where $\norm{\cdot}$ is an appropriate norm.
Typically, box constraints, \ie $\norm{\cdot}_{\infty}$, are used. %
Note that since $x_{k}$ is derived from $u_{i}$ and $y_{i}$ with $i < k$, the trust region is defined on $u_{k}$ and $\tilde{y}_{k}$.
The trust region size $\rho$ is adapted by the sequential convex programming algorithm presented in Section~\ref{sec:lingp-mpc-algorithm}. 

Putting everything together, the complete local problem, which we will call linGP-MPC, is given in \eqref{eq:lingpmpc} below.
\begin{subequations}
\label{eq:lingpmpc}
\begin{align}
  & \text{\bfseries (linGP-MPC)} \nonumber\\
  & \underset{\Delta \UUU_{t}, \ZZZ_{t}}{\text{minimize}} \quad J\left(\tilde{\YYY}_{t}, \tilde{\SSS}_{y,t}, \UUU^{\star}_{t} + \Delta\UUU_{t}, \ZZZ_{t}\right) \label{eq:lingpmpc:obj}\\
  & \text{subject to} \nonumber \\
  & \quad \tilde{y}_{k} = \hat{m}_{x^{\star}_{k}, u^{\star}_{k}}^{T} \xi_{k},  %
    \quad \tilde{\sigma}_{y,k}^{2} = \xi_{k}^{T} \hat{V}_{x^{\star}_{k}, u^{\star}_{k}} \xi_{k} \label{eq:lingpmpc:lingp}\\  %
  & \quad \xi_{k}^{T} = \left[ 1, \Delta x_{k}^{T}, \Delta u_{k}^{T}\right] \\
  & \quad \Delta \tilde{y}_{k} = \tilde{y}_{k} - \overbar{y}_{k}^{\star}, \quad
    \text{$\Delta x_{k}$ defined in \eqref{eq:diff-state-vector}}  \\
  & \quad \norm{\Delta u_{k}} \leq \rho, \quad \norm{\Delta \tilde{y}_{k}} \leq \rho \label{eq:lingpmpc:trustregion} \\
  & \quad z_{k} \in Z, \quad u^{\star}_{k}+\Delta u_{k} \in U \label{eq:lingpmpc:varconstraints}\\
  & \quad g_{i} \left(\tilde{\YYY}_{t}, \tilde{\SSS}_{y,t}, \UUU^{\star}_{t} + \Delta\UUU_{t}, \ZZZ_{t}\right) \leq 0, \, i = 1,\dots,n_{\mathrm{ieq}} \label{eq:lingpmpc:ineqs}\\
  & \quad h_{i} \left(\tilde{\YYY}_{t}, \tilde{\SSS}_{y,t}, \UUU^{\star}_{t} + \Delta\UUU_{t}, \ZZZ_{t}\right) = 0, \, i = 1,\dots,n_{\mathrm{eq}} \label{eq:lingpmpc:eqs}
\end{align}
\end{subequations}
where
$k\in\III_{t}$, %
$\tilde{\YYY}_{t}$ and $\tilde{\SSS}_{y,t}$ respectively collect $\tilde{y}_{k}$ and $\tilde{\sigma}^{2}_{y,k}$.

By Assumption~\ref{assumption:convexity}, the linGP-MPC problem \eqref{eq:lingpmpc} is convex %
and therefore can be solved efficiently with a suitable solver.
By sequentially linearizing the GP with \linGP, solving the convex linGP-MPC problem \eqref{eq:lingpmpc}, and updating the nominal solution and trust region appropriately, the non-convex GP-MPC problem can be solved.
The next subsection will present such a sequential convex programming (SCP) algorithm.

\subsection{Sequential convex programming for linGP-MPC}
\label{sec:lingp-mpc-algorithm}

An issue that can arise in the SCP approach is that the local linGP-MPC problem \eqref{eq:lingpmpc} can be infeasible even though the non-convex GP-MPC problem \eqref{eq:gpmpc} is feasible.
The nominal control inputs $u^{\star}_{k}$ are assumed to satisfy the control constraints \eqref{eq:lingpmpc:varconstraints}: $u^{\star}_{k} \in U$; otherwise we can project them onto $U$.
However, the nominal values may not be feasible for the constraints \eqref{eq:lingpmpc:ineqs} and \eqref{eq:lingpmpc:eqs} and, with a trust region bound $\rho$ small enough, %
problem~\eqref{eq:lingpmpc} is infeasible.
This issue often occurs during the early iterations of the SCP algorithm.

To address this issue, we will use \emph{exact penalty formulations} of the GP-MPC and linGP-MPC problems.
A penalized cost function is defined as
\begin{multline}
  \label{eq:gpmpc-penalized-cost}
  \phi\left(\overbar{\YYY}_{t}, \SSS_{y,t}, \UUU_{t}, \ZZZ_{t}\right) = 
  J\left(\overbar{\YYY}_{t}, \SSS_{y,t}, \UUU_{t}, \ZZZ_{t}\right) + \\
  \lambda \big(
    \textstyle\sum_{i=1}^{n_{\mathrm{ieq}}} \max\left( 0, g_{i} \left(\overbar{\YYY}_{t}, \SSS_{y,t}, \UUU_{t}, \ZZZ_{t}\right)\right) + \\
    \textstyle\sum_{i=1}^{n_{\mathrm{eq}}} \abs{h_{i} \left(\overbar{\YYY}_{t}, \SSS_{y,t}, \UUU_{t}, \ZZZ_{t}\right)}
  \big)
\end{multline}
where $\lambda > 0$ is the penalty weight.
If $\lambda$ is sufficiently large, the original GP-MPC problem can be solved by minimizing $\phi$ subject to the constraints \eqref{eq:gpmpc:mean} to \eqref{eq:gpmpc:varconstraints}.
The exact penalty formulation of the linGP-MPC problem is
\begin{align}
  & \text{\bfseries (linGP-MPC exact penalty)} \nonumber\\
  & \underset{\Delta \UUU_{t}, \ZZZ_{t}}{\text{minimize}} \quad \phi\big(\tilde{\YYY}_{t}, \tilde{\SSS}_{y,t}, \UUU^{\star}_{t} + \Delta\UUU_{t}, \ZZZ_{t}\big) \label{eq:lingpmpc-penalized}\\
  & \text{subject to constraints \eqref{eq:lingpmpc:lingp} to \eqref{eq:lingpmpc:varconstraints}} \text. \nonumber
\end{align}
Note that while both exact penalty formulations use the penalized cost function $\phi$ in \eqref{eq:gpmpc-penalized-cost}, $\phi$ is evaluated at the exact GP outputs ($\overbar{\YYY}_{t}$ and $\SSS_{y,t}$) in GP-MPC, whereas in linGP-MPC, it is evaluated at the approximate \linGP outputs ($\tilde{\YYY}_{t}$ and $\tilde{\SSS}_{y,t}$).

We now present the linGP-SCP algorithm, inspired by the successive convexification algorithm in \cite{mao2016Successiveconvexificationnonconvex}.
At iteration $j$, for $k \in \III_{t}$, let $u_{k}^{(j)}$ and $z_{k}^{(j)}$ be the current solution, $\overbar{y}_{k}^{(j)}$ and $x_{k}^{(j)}$ the corresponding outputs and states, $\rho^{(j)}$ the current trust region size, and $\phi^{(j)}$ the current exact penalized cost.
We obtain the linGP model \linGPModel{f}{x_{k}^{(j)}, u_{k}^{(j)}}, specifically the vectors $\hat{m}_{{x_{k}^{(j)}, u_{k}^{(j)}}}$ and the SDP matrices $\hat{V}_{{x_{k}^{(j)}, u_{k}^{(j)}}}$, for each $k$.
The convex problem \eqref{eq:lingpmpc-penalized} is then solved, resulting in the approximate solution $(\tilde{\YYY}_{t}, \tilde{\SSS}_{y,t}, \tilde{\UUU}_{t} = \UUU_{t}^{(j)} + \Delta \tilde{\UUU}_{t}, \tilde{\ZZZ}_{k})$.
By simulating the GP model \GPModel[f] with the inputs $\tilde{\UUU}_{t}$, we obtain the exact output mean $\overbar{\YYY}_{t}$ and variance $\SSS_{y,t}$.
To judge the algorithm progress, or the quality of the \linGP approximation used in linGP-MPC, we compare the \emph{actual} cost reduction
\begin{equation}
  \label{eq:cost-reduction-actual}
  \delta^{(j)} = \phi^{(j)} - \phi\big(\overbar{\YYY}_{t}, \SSS_{y,t}, \tilde{\UUU}_{t}, \tilde{\ZZZ}_{t}\big)
\end{equation}
to the \emph{predicted} cost reduction
\begin{equation}
  \label{eq:cost-reduction-predicted}
  \tilde{\delta}^{(j)} = \phi^{(j)} - \phi\big(\tilde{\YYY}_{t}, \tilde{\SSS}_{y,t}, \tilde{\UUU}_{t}, \tilde{\ZZZ}_{t}\big) \text.
\end{equation}
If $|\tilde{\delta}^{(j)}| \leq \epsilon$, where $\epsilon > 0$ is a predefined tolerance, the solution is considered converged and the algorithm is terminated.
Otherwise, the ratio $r^{(j)} = \delta^{(j)} / \tilde{\delta}^{(j)}$ is inspected.
Given three predefined thresholds $0 < r_{0} < r_{1} < r_{2} < 1$, there are four possibilities:
\begin{enumerate}
\item $r^{(j)} < r_{0}$: the approximation is considered too inaccurate, hence the approximate solution is rejected and the trust region size $\rho^{(j)}$ is contracted by a predefined factor $\beta^{\mathrm{fail}} < 1$;
\item $r_{0} \leq r^{(j)} < r_{1}$: the approximation is deemed inaccurate but acceptable, hence the approximate solution is accepted but the trust region size $\rho^{(j)}$ is still contracted by $\beta^{\mathrm{fail}}$;
\item $r_{1} \leq r^{(j)} < r_{2}$: the approximation is considered sufficiently accurate, hence the approximate solution is accepted and the trust region size $\rho^{(j)}$ is retained;
\item $r^{(j)} \geq r_{2}$: the approximation is deemed highly accurate or even conservative, hence the approximate solution is accepted and the trust region size $\rho^{(j)}$ is enlarged by a predefined factor $\beta^{\mathrm{succ}} > 1$.
\end{enumerate}
The iteration is repeated until convergence or until a maximum number of iterations $j_{\mathrm{max}}>0$ is reached.
The above linGP-SCP algorithm is detailed in Algorithm~\ref{alg:lingp-mpc-scp}.

\begin{algorithm}
\caption{linGP-SCP: Sequential Convex Programming for linGP-MPC Problem}
\label{alg:lingp-mpc-scp}
\begin{algorithmic}[1]
  \Require %
  $\UUU_{t}^{(0)} = \{u_{k}^{(0)} \in U \;|\; k \in \III_{t}\}$, $\ZZZ_{t}^{(0)} = \{z_{k}^{(0)} \in Z \;|\; k \in \III_{t}\}$, %
  $\rho^{(0)} > 0$, $\lambda > 0$, $0 < r_{0} < r_{1} < r_{2} < 1$, $\beta^{\mathrm{fail}}<1$, $\beta^{\mathrm{succ}}>1$, $\epsilon > 0$, $j_{\mathrm{max}} > 0$
  \State Simulate \GPModel[f] with $\UUU_{t}^{(0)}$, obtain $\overbar{\YYY}_{t}^{(0)}$, $\SSS_{y,t}^{(0)}$ %
  \State $\phi^{(0)} \leftarrow \phi\big(\overbar{\YYY}_{t}^{(0)}, \SSS_{y,t}^{(0)}, \UUU_{t}^{(0)}, \ZZZ_{t}^{(0)}\big)$
  \For{$j = 0, \dots, j_{\mathrm{max}}-1$}
    \State Compute $\hat{m}_{{x_{k}^{(j)}, u_{k}^{(j)}}}$ and $\hat{V}_{{x_{k}^{(j)}, u_{k}^{(j)}}}$  $\forall k\in\III_{t}$ by \eqref{eq:derivative-distribution}
    \State Solve problem~\eqref{eq:lingpmpc-penalized} to get $\tilde{\YYY}_{t}$, $\tilde{\SSS}_{y,t}$, $\tilde{\UUU}_{t}$, $\tilde{\ZZZ}_{k}$
    \State Simulate \GPModel[f] with $\tilde{\UUU}_{t}$, obtain $\overbar{\YYY}_{t}$, $\SSS_{y,t}$
    \State $\delta^{(j)} \leftarrow \phi^{(j)} - \phi\big(\overbar{\YYY}_{t}, \SSS_{y,t}, \tilde{\UUU}_{t}, \tilde{\ZZZ}_{t}\big)$
    \State $\tilde{\delta}^{(j)} \leftarrow \phi^{(j)} - \phi\big(\tilde{\YYY}_{t}, \tilde{\SSS}_{y,t}, \tilde{\UUU}_{t}, \tilde{\ZZZ}_{t}\big)$
    \If{$|\tilde{\delta}^{(j)}| \leq \epsilon$}
      \State \textbf{stop, return $\UUU_{t}^{(j)}$, $\ZZZ_{t}^{(j)}$}
    \EndIf
    \State $r^{(j)} \leftarrow \delta^{(j)} / \tilde{\delta}^{(j)}$
    \If{$r^{(j)} < r_{0}$}
      \State \multiline{Keep current solution: $\UUU_{t}^{(j+1)} \leftarrow \UUU_{t}^{(j)}$, $\ZZZ_{t}^{(j+1)} \leftarrow \ZZZ_{t}^{(j)}$, $\overbar{\YYY}_{t}^{(j+1)} \leftarrow \overbar{\YYY}_{t}^{(j)}$, $\SSS_{y,t}^{(j+1)} \leftarrow \SSS_{y,t}^{(j)}$}
      \State $\rho^{(j+1)} \leftarrow \beta^{\mathrm{fail}}\rho^{(j)}$ 
    \Else
      \State \multiline{Accept solution: $\UUU_{t}^{(j+1)} \leftarrow \tilde{\UUU}_{t}$, $\ZZZ_{t}^{(j+1)} \leftarrow \tilde{\ZZZ}_{t}$, $\overbar{\YYY}_{t}^{(j+1)} \leftarrow \overbar{\YYY}_{t}$, $\SSS_{y,t}^{(j+1)} \leftarrow \SSS_{y,t}$}
      \If{$r^{(j)} < r_{1}$}
        \State $\rho^{(j+1)} \leftarrow \beta^{\mathrm{fail}}\rho^{(j)}$
      \ElsIf{$r^{(j)} < r_{2}$}
        \State $\rho^{(j+1)} \leftarrow \rho^{(j)}$
      \Else
        \State $\rho^{(j+1)} \leftarrow \beta^{\mathrm{succ}}\rho^{(j)}$ 
      \EndIf
    \EndIf
  \EndFor  %
  \State \textbf{return $\UUU_{t}^{(j_{\mathrm{max}})}$, $\ZZZ_{t}^{(j_{\mathrm{max}})}$}  
\end{algorithmic}
\end{algorithm}

\subsection{Complexity of linGP-SCP algorithm}
\label{sec:complexity-lingp-mpc}

The linGP-SCP algorithm~\ref{alg:lingp-mpc-scp} consists of a main loop and a convex local subproblem (\ie the linGP-MPC problem).
Because a general MPC formulation is assumed, resulting in a general convex subproblem, it is not possible to determine the specific complexity of the subproblem.
However, we note that the convex subproblem can usually be solved efficiently.
In this subsection, we focus instead on the complexity of the algorithm with respect to the size of the GP model (\ie the size $N$ of its training data).

An important feature of the linGP-SCP algorithm is that the subproblem's complexity is independent of the size of the GP model.
This is obvious from inspecting the linGP-MPC subproblem in \eqref{eq:lingpmpc}, where the parameters $\hat{m}$ and $\hat{V}$ do not depend on $N$.
Indeed, the numerical example in Section~\ref{sec:simulation} will show that the time spent on solving the subproblem is steady regardless of the size $N$ of the GP model.
As a consequence, the \emph{inner iterations} of the linGP-SCP algorithm, which solve the convex subproblem in each \emph{outer iteration} of the main loop, do not depend on $N$.

The main loop involves, among other computations, the simulation and linearization of the GP model, which scale cubically with $N$.
In fact, for a sufficiently large GP model, $N$ is the largest factor determining the complexity and the solving time of the GP-MPC problem.
Therefore, we need to consider the number of iterations typically required by the linGP-SCP algorithm.
Unlike standard trust-region algorithms, which usually perform a line search to achieve cost reduction, the linGP-SCP algorithm optimally solves the convex subproblem to reduce the cost in each main succession.
Hence, the number of inner iterations in each outer iteration is increased, while the number of outer iterations is decreased.
In other words, by achieving a greater cost reduction in each outer iteration, at the expense of increased subproblem solving time, the linGP-SCP algorithm can reduce the number of outer iterations.
Because the inner iterations are independent of $N$ while the outer iterations scale cubically with $N$, this feature of the linGP-SCP algorithm helps reduce the overall complexity and solving time significantly, particularly for large GP models (\ie large $N$).

Compared to SQP-based methods, the linGP-SCP algorithm has a crucial advantage, stemming from the use of linGP models to approximate the original GP model.
An SQP-based method typically uses a computationally expensive technique such as the BFGS update to estimate the Hessian of the cost function and probably other functions.
Moreover, because the exact GP mean and variance, as computed by \eqref{eq:gp-regression}, are used directly, their Hessians need to be computed, whose computational complexity grow exponentially with $N$.
By contrast, the linGP approximation, as expressed in \eqref{eq:derivative-distribution} and \eqref{eq:lingp-mean-var}, relies only on the Jacobians of the covariance function and therfore is much cheaper to compute.
Another distinction between the linGP-SCP algorithm and SQP-based methods is that the linGP approximation, as a valid GP in itself, always guarantees the positive definiteness of its variance (\cf \eqref{eq:gp-linearized-final} and the paragraph following it).
Whereas there is no such guarantee in the direct approximation of the GP variance used by SQP-based methods.
Therefore, SQP-based methods must employ additional measures to ensure the positive definiteness of the estimated Hessian so that the SQP subproblem is convex, incurring more computational cost.

In summary, the use of the linGP approximation, with first-order computation and guaranteed convexity, and its resulting convex linGP-MPC problem in the linGP-SCP algorithm leads to a significant reduction in computational complexity compared to other methods.
This advantage is demonstrated in the numerical example in the next section.

%% file: simulation.tex
To demonstrate the efficacy of the linGP-SCP algorithm, we considered a GP-MPC problem of a nonlinear dynamic system, adapted from \cite[Example~4.3, p 183]{Kocijan2016}.
The process to be controlled is a discrete-time scalar system with noisy output of the form:
\begin{equation}
  \label{eq:simulation:dynamics}
  x_{k+1} = x_{k} - 0.5 \tanh(x_{k} + u_{k}^3), \quad y_{k} = x_{k} + w_{k}
\end{equation}
where $w_{k}$ is a white noise with standard deviation $0.025$ and zero mean.
The following input %
constraints are imposed
\begin{equation}
  \label{eq:simulation:u-constraints}
  -1 \leq u_{k} \leq 1, \quad -0.5 \leq u_{k+1} - u_{k} \leq 0.5, \quad \forall k \text.
\end{equation}

This nonlinear process was learned by GP models using the autoregressive input $[y_{k-1}, u_{k}]$ and the standard squared exponential covariance function (see \cite{Kocijan2016,Rasmussen2006}).
The training data were obtained by feeding a random input signal that satisfies the constraints~\eqref{eq:simulation:u-constraints} to the process \cite{Kocijan2016}.
To illustrate the performance of solving the GP-MPC problem with respect to the training data size $N$, we trained 15 GP models with $N$ incremented by 100 from 100 to 1,500.
It is worth pointing out that a training data set larger than $N=400$ did not improve the GP model accuracy significantly; however, we considered large values of $N$ to illustrate the computational benefits of our algorithm.

The control goal is to track a given reference signal $r_{k}$.
The GP-MPC problem is formulated as follows \cite{Kocijan2016}
\begin{subequations}
\label{eq:simulation:gpmpc}
\begin{align}
  & \underset{u_{t}, \dots, u_{t+H-1}}{\text{minimize}} \; \textstyle\sum_{k=t}^{t+H-1} \left( Q (\overbar{y}_{k} - r_{t})^{2} + R (u_{k}-u_{k-1})^{2} \right) \nonumber \\%\label{eq:simulation:gpmpc:obj}\\
  & \text{subject to} \nonumber \\
  & \quad \overbar{y}_{k} = \GPmean{f}(\overbar{y}_{k-1}, u_{k}), \quad \sigma_{y,k}^{2} = \GPvar{f}(\overbar{y}_{k-1}, u_{k}) \\  %
  & \quad -1 \leq u_{k} \leq 1, \quad -0.5 \leq u_{k+1} - u_{k} \leq 0.5 \\
  & \quad -1.2 \leq \overbar{y}_{k} - 2 \sigma_{y,k}, \quad \overbar{y}_{k} + 2 \sigma_{y,k} \leq 1.2 \label{eq:simulation:gpmpc:state-constraint}\\
  & \quad -\delta \leq \overbar{y}_{t+H-1} - r_{t} - 2 \sigma_{y,t+H-1} \label{eq:simulation:gpmpc:terminal-constraint-l}\\
  & \quad \overbar{y}_{t+H-1} - r_{t} + 2 \sigma_{y,t+H-1} \leq \delta \label{eq:simulation:gpmpc:terminal-constraint-h}
\end{align}
\end{subequations}
The state constraints~\eqref{eq:simulation:gpmpc:state-constraint} and terminal constraints~\eqref{eq:simulation:gpmpc:terminal-constraint-l} and \eqref{eq:simulation:gpmpc:terminal-constraint-h} are deterministic approximations of stochastic constraints that, with high probability, bound the process state.
The MPC parameters are $H=12$, $Q=10$, $R=0.1$, and $\delta = 0.075$.
The closed-loop control system was simulated for 100 steps with the following reference signal:
\[r_k = -0.5, \; k \in [0, 50]; \quad r_k = -0.2, \; k \in [51, 100]\text.\]

All simulations were run in Matlab R2017b on an iMac with Intel Core i7 4.2 GHz processor and 16 Gb RAM.

\subsection{Solving the GP-MPC}
\label{sec:simulation:solving}

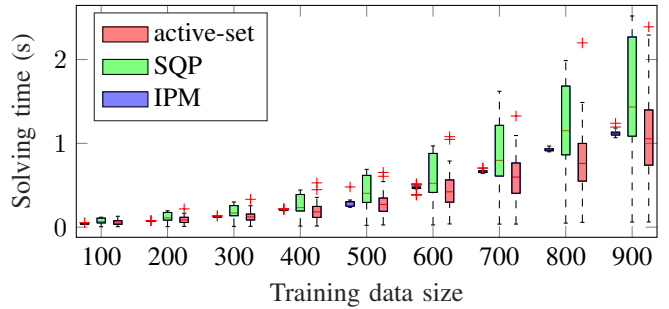
\begin{figure}[!t]
  \centering
  \setlength{\figwidth}{0.94\columnwidth}
  \setlength{\figheight}{0.38\figwidth}
  \input{figs/sqp_vs_ip}
  \caption{%
    Solving time of different NLP methods for the example in Section~\ref{sec:simulation}.}
  \label{fig:sqp-performance}
\end{figure}

Figure~\ref{fig:sqp-performance} shows %
a box plot of the solving time of the GP-MPC problem \eqref{eq:simulation:gpmpc} for increasing training data size and with three NLP methods: IPM, SQP, and active-set.
In all cases, the solving time grows quickly with the training data size.
While the IPM solving time does not vary much between time steps (\ie low variance), the SQP and active-set methods have noticeably varying solving time (\ie high variance with far-away outliers).
In other words, the performance of IPM is steady though slow, while SQP and active-set methods are unpredictable in terms of performance.
For these reasons, general NLP methods %
are not well suited for real-time GP-MPC applications, where the optimization solving performance must be both fast and consistent.

To compare the performance of the \linGP-SCP algorithm with state-of-the-art solvers, we solved the GP-MPC problem \eqref{eq:simulation:gpmpc}, for all 15 GP models, with different solvers:
\begin{itemize}
\item \textbf{Ipopt}: the GP-MPC was formulated in CasADi \cite{Andersson2018} for Matlab and solved by the open-source solver Ipopt \cite{Waechter2009b} with the optimized linear solver MA57.  All Jacobians and Hessians were calculated automatically by CasADi in C code, well-known to be very fast.
\item \textbf{Knitro}: similar to the previous, however the commercial nonlinear solver Knitro %
  was used.  Knitro is commonly regarded as one of the best nonlinear solvers.
\item \textbf{linGP-SCP}: the linGP-SCP algorithm was prototyped purely in Matlab, where the subproblem was modeled by Yalmip \cite{Yalmip2004} and solved by Gurobi. %
\end{itemize}
The solvers were configured to similar accuracy levels; indeed, given the same parameters, their solutions %
were always within \num{1e-6} from each other.
We make two remarks.
\begin{remark}
  We also used the commercial solver \textbf{fmincon}, included in Matlab, with its SQP and active-set methods.
  However, as discussed earlier and illustrated in Figure~\ref{fig:sqp-performance}, its performance was unpredictable and often worse than other solvers.
  Therefore we will not report its results here.
\end{remark}
\begin{remark}
  Unlike the highly optimized solvers Ipopt and Knitro, which are usually implemented in C/C++ and/or Fortran, the linGP-SCP prototype in Matlab is certainly not an optimized implementation.
  In addition, the Yalmip toolbox %
  is known to be slower than the C-based CasADi library. %
  Therefore, there is significant room to improve the performance of the linGP-SCP implementation.
\end{remark}

\subsection{Simulation results and discussions}
\label{sec:simulation:results}

In all cases, the GP-MPC controller was able to track the given reference.
Because our main interest is in the solving performance, we omit the tracking control results and only report the performance results of the used solvers.

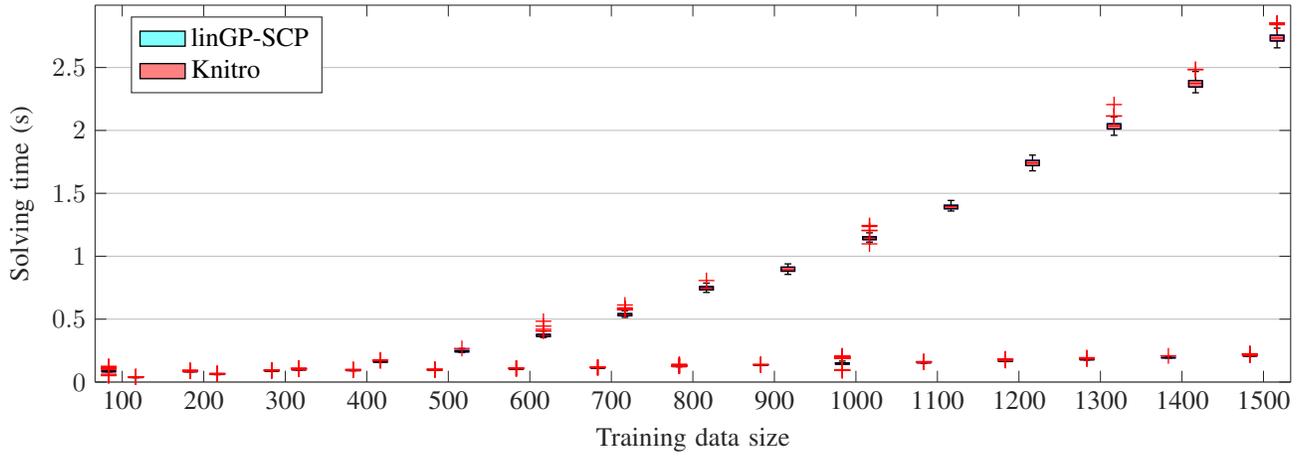
\begin{figure*}[!tb]
  \centering
  \setlength{\figwidth}{0.94\textwidth}
  \setlength{\figheight}{0.3\figwidth}
  \input{figs/solvingtime}
  \caption{Box plot of solving time of each MPC step by Knitro and linGP-SCP, for all GP model sizes.}
  \label{fig:solving-time}
\end{figure*}

Figure~\ref{fig:solving-time} is a box plot of the solving time of each MPC step by Knitro and linGP-SCP, for all 15 GP models.
We do not report the timing results of Ipopt in Figure~\ref{fig:solving-time} because they were significantly worse than the other two.
Observe that linGP-SCP's solving time was highly consistent with low variance and few small outliers.

\textbf{Most importantly}, while the solving time of %
Knitro grew quickly with the GP model size $N$, the solving time of linGP-SCP grew much slower. %
This observation was more obvious in Figure~\ref{fig:solving-time-median}, which plots the median solving time.
Clearly, the solving time of Ipopt and Knitro grew exponentially with $N$ while the solving time of linGP-SCP did not increase significantly.
Figure~\ref{fig:solving-time-median-ratio} shows the relative ratio of the median solving time of Ipopt and Knitro to that of linGP-SCP.
It can be seen that Ipopt and Knitro became increasingly slower than linGP-SCP when $N$ increased.
These results confirm the complexity analysis in Section~\ref{sec:complexity-lingp-mpc} that, compared to other methods, the complexity of linGP-SCP is less affected by the GP model size and therefore grows slowly with $N$.
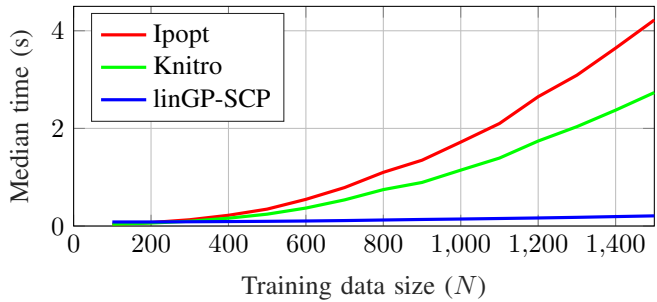
\begin{figure}[!tb]
  \centering
  \setlength{\figwidth}{0.94\columnwidth}
  \setlength{\figheight}{0.36\figwidth}
  \input{figs/solvingtime_median}
  \caption{Median solving time trend of the three solvers.}
  \label{fig:solving-time-median}
\end{figure}
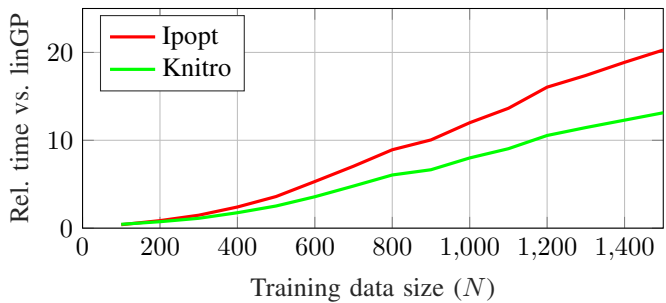
\begin{figure}[!tb]
  \centering
  \setlength{\figwidth}{0.94\columnwidth}
  \setlength{\figheight}{0.36\figwidth}
  \input{figs/solvingtime_median_ratio}
  \caption{Relative median time versus linGP-SCP.}
  \label{fig:solving-time-median-ratio}
\end{figure}

In Table~\ref{tbl:simulation:solving-time}, the mean and standard deviation of the execution time (in seconds) per MPC step for several model sizes are listed.
Again, these numbers support the previous claim about the performance advantage of the linGP-SCP algorithm.
We can also observe that the solving time of the linGP-MPC subproblem (the ``linGP-inner'' row in the table) was constant regardless of the GP model size, as stated in the complexity analysis in Section~\ref{sec:complexity-lingp-mpc}.

\begin{table*}[!tb]
  \centering
  \begin{tabular}{lllllllllllll}
    \toprule
    & \multicolumn{2}{c}{$N=500$}
    & \multicolumn{2}{c}{$N=700$}
    & \multicolumn{2}{c}{$N=900$}
    & \multicolumn{2}{c}{$N=1100$}
    & \multicolumn{2}{c}{$N=1300$}
    & \multicolumn{2}{c}{$N=1500$} \\
    
    \cmidrule(r){2-3}
    \cmidrule(r){4-5}
    \cmidrule(r){6-7}
    \cmidrule(r){8-9}
    \cmidrule(r){10-11}
    \cmidrule(r){12-13}
    
    & $\mu$ & $\sigma$ & $\mu$ & $\sigma$ & $\mu$ & $\sigma$ & $\mu$ & $\sigma$ & $\mu$ & $\sigma$ & $\mu$ & $\sigma$ \\
    \midrule
    linGP & 0.096 & 0.0011 & 0.112 & 0.0015 & 0.134 & 0.0013 & 0.154 & 0.0020 & 0.178 & 0.00291 & 0.209 & 0.0042 \\
    linGP-inner & 0.024 & 0.0002 & 0.024 & 0.0003 & 0.024 & 0.0002 & 0.024 & 0.0003 & 0.024 & 0.0002 & 0.024 & 0.0003 \\
    Knitro & 0.246 & 0.0081 & 0.538 & 0.0172 & 0.896 & 0.0208 & 1.390 & 0.0175 & 2.040 & 0.0365 & 2.740 & 0.0390 \\
    Ipopt & 0.331 & 0.0385 & 0.751 & 0.0909 & 1.29 & 0.1700 & 2.010 & 0.2580 & 2.950 & 0.3680 & 4.010 & 0.4680 \\
    \bottomrule
  \end{tabular}
  \caption{Solving time statistics for several GP model sizes.}
  \label{tbl:simulation:solving-time}
\end{table*}

%% file: figs/sqp_vs_ip.tex
\begin{tikzpicture}

\begin{axis}[%
width=0.951\figwidth,
height=\figheight,
at={(0\figwidth,0\figheight)},
scale only axis,
unbounded coords=jump,
xmin=1.125,
xmax=9.875,
xtick={1.5,2.5,3.5,4.5,5.5,6.5,7.5,8.5,9.5},
xticklabels={{100},{200},{300},{400},{500},{600},{700},{800},{900}},
xlabel style={font=\color{white!15!black}},
xlabel={Training data size},
ymin=-0.12079867835,
ymax=2.64523132135,
ylabel style={font=\color{white!15!black}},
ylabel={Solving time (s)},
axis background/.style={fill=white},
legend style={at={(0.03,0.97)}, anchor=north west, legend cell align=left, align=left, draw=white!15!black}
]
\addplot [color=black, dashed, forget plot]
  table[row sep=crcr]{%
1.25	0.04068242\\
1.25	0.044908605\\
};
\addplot [color=black, dashed, forget plot]
  table[row sep=crcr]{%
1.5	0.107235153\\
1.5	0.118589241\\
};
\addplot [color=black, dashed, forget plot]
  table[row sep=crcr]{%
1.75	0.076390559\\
1.75	0.129793265\\
};
\addplot [color=black, dashed, forget plot]
  table[row sep=crcr]{%
2.25	0.073786595\\
2.25	0.075943542\\
};
\addplot [color=black, dashed, forget plot]
  table[row sep=crcr]{%
2.5	0.1752357035\\
2.5	0.195409685\\
};
\addplot [color=black, dashed, forget plot]
  table[row sep=crcr]{%
2.75	0.1175657075\\
2.75	0.16529358\\
};
\addplot [color=black, dashed, forget plot]
  table[row sep=crcr]{%
3.25	0.1243548345\\
3.25	0.129525739\\
};
\addplot [color=black, dashed, forget plot]
  table[row sep=crcr]{%
3.5	0.2574689365\\
3.5	0.300760458\\
};
\addplot [color=black, dashed, forget plot]
  table[row sep=crcr]{%
3.75	0.156054798\\
3.75	0.25484652\\
};
\addplot [color=black, dashed, forget plot]
  table[row sep=crcr]{%
4.25	0.20554893375\\
4.25	0.212773562\\
};
\addplot [color=black, dashed, forget plot]
  table[row sep=crcr]{%
4.5	0.3897081445\\
4.5	0.442956114\\
};
\addplot [color=black, dashed, forget plot]
  table[row sep=crcr]{%
4.75	0.247085482\\
4.75	0.355352751\\
};
\addplot [color=black, dashed, forget plot]
  table[row sep=crcr]{%
5.25	0.3050969475\\
5.25	0.324048885\\
};
\addplot [color=black, dashed, forget plot]
  table[row sep=crcr]{%
5.5	0.617178367\\
5.5	0.689724464\\
};
\addplot [color=black, dashed, forget plot]
  table[row sep=crcr]{%
5.75	0.3471068555\\
5.75	0.54413642\\
};
\addplot [color=black, dashed, forget plot]
  table[row sep=crcr]{%
6.25	0.47848464025\\
6.25	0.500870818\\
};
\addplot [color=black, dashed, forget plot]
  table[row sep=crcr]{%
6.5	0.87937192\\
6.5	0.969422913\\
};
\addplot [color=black, dashed, forget plot]
  table[row sep=crcr]{%
6.75	0.563423231\\
6.75	0.789278813\\
};
\addplot [color=black, dashed, forget plot]
  table[row sep=crcr]{%
7.25	0.671889845\\
7.25	0.69795132\\
};
\addplot [color=black, dashed, forget plot]
  table[row sep=crcr]{%
7.5	1.2144539735\\
7.5	1.621332047\\
};
\addplot [color=black, dashed, forget plot]
  table[row sep=crcr]{%
7.75	0.7657274065\\
7.75	1.093049018\\
};
\addplot [color=black, dashed, forget plot]
  table[row sep=crcr]{%
8.25	0.935355293\\
8.25	0.967088953\\
};
\addplot [color=black, dashed, forget plot]
  table[row sep=crcr]{%
8.5	1.68311927\\
8.5	1.98731626\\
};
\addplot [color=black, dashed, forget plot]
  table[row sep=crcr]{%
8.75	0.9987906975\\
8.75	1.487425632\\
};
\addplot [color=black, dashed, forget plot]
  table[row sep=crcr]{%
9.25	1.132957689\\
9.25	1.182381003\\
};
\addplot [color=black, dashed, forget plot]
  table[row sep=crcr]{%
9.5	2.26860556\\
9.5	2.519502685\\
};
\addplot [color=black, dashed, forget plot]
  table[row sep=crcr]{%
9.75	1.3982497295\\
9.75	2.29151871\\
};
\addplot [color=black, dashed, forget plot]
  table[row sep=crcr]{%
1.25	0.034485278\\
1.25	0.0367141805\\
};
\addplot [color=black, dashed, forget plot]
  table[row sep=crcr]{%
1.5	0.004929958\\
1.5	0.050791497\\
};
\addplot [color=black, dashed, forget plot]
  table[row sep=crcr]{%
1.75	0.005346137\\
1.75	0.0341194575\\
};
\addplot [color=black, dashed, forget plot]
  table[row sep=crcr]{%
2.25	0.070550918\\
2.25	0.072214465\\
};
\addplot [color=black, dashed, forget plot]
  table[row sep=crcr]{%
2.5	0.006869534\\
2.5	0.0841946085\\
};
\addplot [color=black, dashed, forget plot]
  table[row sep=crcr]{%
2.75	0.00922007\\
2.75	0.0557341925\\
};
\addplot [color=black, dashed, forget plot]
  table[row sep=crcr]{%
3.25	0.118304568\\
3.25	0.120835236\\
};
\addplot [color=black, dashed, forget plot]
  table[row sep=crcr]{%
3.5	0.009501108\\
3.5	0.1363445055\\
};
\addplot [color=black, dashed, forget plot]
  table[row sep=crcr]{%
3.75	0.010024412\\
3.75	0.0858501055\\
};
\addplot [color=black, dashed, forget plot]
  table[row sep=crcr]{%
4.25	0.197250752\\
4.25	0.20060523175\\
};
\addplot [color=black, dashed, forget plot]
  table[row sep=crcr]{%
4.5	0.013189377\\
4.5	0.193567634\\
};
\addplot [color=black, dashed, forget plot]
  table[row sep=crcr]{%
4.75	0.013436693\\
4.75	0.1160456535\\
};
\addplot [color=black, dashed, forget plot]
  table[row sep=crcr]{%
5.25	0.236556152\\
5.25	0.25028277825\\
};
\addplot [color=black, dashed, forget plot]
  table[row sep=crcr]{%
5.5	0.021124744\\
5.5	0.295282608\\
};
\addplot [color=black, dashed, forget plot]
  table[row sep=crcr]{%
5.75	0.026181225\\
5.75	0.1890805815\\
};
\addplot [color=black, dashed, forget plot]
  table[row sep=crcr]{%
6.25	0.456988268\\
6.25	0.46330008025\\
};
\addplot [color=black, dashed, forget plot]
  table[row sep=crcr]{%
6.5	0.02667143\\
6.5	0.414276705\\
};
\addplot [color=black, dashed, forget plot]
  table[row sep=crcr]{%
6.75	0.03843973\\
6.75	0.2986842225\\
};
\addplot [color=black, dashed, forget plot]
  table[row sep=crcr]{%
7.25	0.641575504\\
7.25	0.651586804\\
};
\addplot [color=black, dashed, forget plot]
  table[row sep=crcr]{%
7.5	0.036023653\\
7.5	0.6095816415\\
};
\addplot [color=black, dashed, forget plot]
  table[row sep=crcr]{%
7.75	0.036707765\\
7.75	0.4037924045\\
};
\addplot [color=black, dashed, forget plot]
  table[row sep=crcr]{%
8.25	0.898771232\\
8.25	0.911693595\\
};
\addplot [color=black, dashed, forget plot]
  table[row sep=crcr]{%
8.5	0.048344332\\
8.5	0.863271304\\
};
\addplot [color=black, dashed, forget plot]
  table[row sep=crcr]{%
8.75	0.057096286\\
8.75	0.5489324535\\
};
\addplot [color=black, dashed, forget plot]
  table[row sep=crcr]{%
9.25	1.067073728\\
9.25	1.0988397155\\
};
\addplot [color=black, dashed, forget plot]
  table[row sep=crcr]{%
9.5	0.060576498\\
9.5	1.084868247\\
};
\addplot [color=black, dashed, forget plot]
  table[row sep=crcr]{%
9.75	0.062514688\\
9.75	0.7394289345\\
};
\addplot [color=black, forget plot]
  table[row sep=crcr]{%
1.21875	0.044908605\\
1.28125	0.044908605\\
};
\addplot [color=black, forget plot]
  table[row sep=crcr]{%
1.46875	0.118589241\\
1.53125	0.118589241\\
};
\addplot [color=black, forget plot]
  table[row sep=crcr]{%
1.71875	0.129793265\\
1.78125	0.129793265\\
};
\addplot [color=black, forget plot]
  table[row sep=crcr]{%
2.21875	0.075943542\\
2.28125	0.075943542\\
};
\addplot [color=black, forget plot]
  table[row sep=crcr]{%
2.46875	0.195409685\\
2.53125	0.195409685\\
};
\addplot [color=black, forget plot]
  table[row sep=crcr]{%
2.71875	0.16529358\\
2.78125	0.16529358\\
};
\addplot [color=black, forget plot]
  table[row sep=crcr]{%
3.21875	0.129525739\\
3.28125	0.129525739\\
};
\addplot [color=black, forget plot]
  table[row sep=crcr]{%
3.46875	0.300760458\\
3.53125	0.300760458\\
};
\addplot [color=black, forget plot]
  table[row sep=crcr]{%
3.71875	0.25484652\\
3.78125	0.25484652\\
};
\addplot [color=black, forget plot]
  table[row sep=crcr]{%
4.21875	0.212773562\\
4.28125	0.212773562\\
};
\addplot [color=black, forget plot]
  table[row sep=crcr]{%
4.46875	0.442956114\\
4.53125	0.442956114\\
};
\addplot [color=black, forget plot]
  table[row sep=crcr]{%
4.71875	0.355352751\\
4.78125	0.355352751\\
};
\addplot [color=black, forget plot]
  table[row sep=crcr]{%
5.21875	0.324048885\\
5.28125	0.324048885\\
};
\addplot [color=black, forget plot]
  table[row sep=crcr]{%
5.46875	0.689724464\\
5.53125	0.689724464\\
};
\addplot [color=black, forget plot]
  table[row sep=crcr]{%
5.71875	0.54413642\\
5.78125	0.54413642\\
};
\addplot [color=black, forget plot]
  table[row sep=crcr]{%
6.21875	0.500870818\\
6.28125	0.500870818\\
};
\addplot [color=black, forget plot]
  table[row sep=crcr]{%
6.46875	0.969422913\\
6.53125	0.969422913\\
};
\addplot [color=black, forget plot]
  table[row sep=crcr]{%
6.71875	0.789278813\\
6.78125	0.789278813\\
};
\addplot [color=black, forget plot]
  table[row sep=crcr]{%
7.21875	0.69795132\\
7.28125	0.69795132\\
};
\addplot [color=black, forget plot]
  table[row sep=crcr]{%
7.46875	1.621332047\\
7.53125	1.621332047\\
};
\addplot [color=black, forget plot]
  table[row sep=crcr]{%
7.71875	1.093049018\\
7.78125	1.093049018\\
};
\addplot [color=black, forget plot]
  table[row sep=crcr]{%
8.21875	0.967088953\\
8.28125	0.967088953\\
};
\addplot [color=black, forget plot]
  table[row sep=crcr]{%
8.46875	1.98731626\\
8.53125	1.98731626\\
};
\addplot [color=black, forget plot]
  table[row sep=crcr]{%
8.71875	1.487425632\\
8.78125	1.487425632\\
};
\addplot [color=black, forget plot]
  table[row sep=crcr]{%
9.21875	1.182381003\\
9.28125	1.182381003\\
};
\addplot [color=black, forget plot]
  table[row sep=crcr]{%
9.46875	2.519502685\\
9.53125	2.519502685\\
};
\addplot [color=black, forget plot]
  table[row sep=crcr]{%
9.71875	2.29151871\\
9.78125	2.29151871\\
};
\addplot [color=black, forget plot]
  table[row sep=crcr]{%
1.21875	0.034485278\\
1.28125	0.034485278\\
};
\addplot [color=black, forget plot]
  table[row sep=crcr]{%
1.46875	0.004929958\\
1.53125	0.004929958\\
};
\addplot [color=black, forget plot]
  table[row sep=crcr]{%
1.71875	0.005346137\\
1.78125	0.005346137\\
};
\addplot [color=black, forget plot]
  table[row sep=crcr]{%
2.21875	0.070550918\\
2.28125	0.070550918\\
};
\addplot [color=black, forget plot]
  table[row sep=crcr]{%
2.46875	0.006869534\\
2.53125	0.006869534\\
};
\addplot [color=black, forget plot]
  table[row sep=crcr]{%
2.71875	0.00922007\\
2.78125	0.00922007\\
};
\addplot [color=black, forget plot]
  table[row sep=crcr]{%
3.21875	0.118304568\\
3.28125	0.118304568\\
};
\addplot [color=black, forget plot]
  table[row sep=crcr]{%
3.46875	0.009501108\\
3.53125	0.009501108\\
};
\addplot [color=black, forget plot]
  table[row sep=crcr]{%
3.71875	0.010024412\\
3.78125	0.010024412\\
};
\addplot [color=black, forget plot]
  table[row sep=crcr]{%
4.21875	0.197250752\\
4.28125	0.197250752\\
};
\addplot [color=black, forget plot]
  table[row sep=crcr]{%
4.46875	0.013189377\\
4.53125	0.013189377\\
};
\addplot [color=black, forget plot]
  table[row sep=crcr]{%
4.71875	0.013436693\\
4.78125	0.013436693\\
};
\addplot [color=black, forget plot]
  table[row sep=crcr]{%
5.21875	0.236556152\\
5.28125	0.236556152\\
};
\addplot [color=black, forget plot]
  table[row sep=crcr]{%
5.46875	0.021124744\\
5.53125	0.021124744\\
};
\addplot [color=black, forget plot]
  table[row sep=crcr]{%
5.71875	0.026181225\\
5.78125	0.026181225\\
};
\addplot [color=black, forget plot]
  table[row sep=crcr]{%
6.21875	0.456988268\\
6.28125	0.456988268\\
};
\addplot [color=black, forget plot]
  table[row sep=crcr]{%
6.46875	0.02667143\\
6.53125	0.02667143\\
};
\addplot [color=black, forget plot]
  table[row sep=crcr]{%
6.71875	0.03843973\\
6.78125	0.03843973\\
};
\addplot [color=black, forget plot]
  table[row sep=crcr]{%
7.21875	0.641575504\\
7.28125	0.641575504\\
};
\addplot [color=black, forget plot]
  table[row sep=crcr]{%
7.46875	0.036023653\\
7.53125	0.036023653\\
};
\addplot [color=black, forget plot]
  table[row sep=crcr]{%
7.71875	0.036707765\\
7.78125	0.036707765\\
};
\addplot [color=black, forget plot]
  table[row sep=crcr]{%
8.21875	0.898771232\\
8.28125	0.898771232\\
};
\addplot [color=black, forget plot]
  table[row sep=crcr]{%
8.46875	0.048344332\\
8.53125	0.048344332\\
};
\addplot [color=black, forget plot]
  table[row sep=crcr]{%
8.71875	0.057096286\\
8.78125	0.057096286\\
};
\addplot [color=black, forget plot]
  table[row sep=crcr]{%
9.21875	1.067073728\\
9.28125	1.067073728\\
};
\addplot [color=black, forget plot]
  table[row sep=crcr]{%
9.46875	0.060576498\\
9.53125	0.060576498\\
};
\addplot [color=black, forget plot]
  table[row sep=crcr]{%
9.71875	0.062514688\\
9.78125	0.062514688\\
};
\addplot [color=blue, forget plot]
  table[row sep=crcr]{%
1.1875	0.0367141805\\
1.1875	0.04068242\\
1.3125	0.04068242\\
1.3125	0.0367141805\\
1.1875	0.0367141805\\
};
\addplot [color=blue, forget plot]
  table[row sep=crcr]{%
1.4375	0.050791497\\
1.4375	0.107235153\\
1.5625	0.107235153\\
1.5625	0.050791497\\
1.4375	0.050791497\\
};
\addplot [color=blue, forget plot]
  table[row sep=crcr]{%
1.6875	0.0341194575\\
1.6875	0.076390559\\
1.8125	0.076390559\\
1.8125	0.0341194575\\
1.6875	0.0341194575\\
};
\addplot [color=blue, forget plot]
  table[row sep=crcr]{%
2.1875	0.072214465\\
2.1875	0.073786595\\
2.3125	0.073786595\\
2.3125	0.072214465\\
2.1875	0.072214465\\
};
\addplot [color=blue, forget plot]
  table[row sep=crcr]{%
2.4375	0.0841946085\\
2.4375	0.1752357035\\
2.5625	0.1752357035\\
2.5625	0.0841946085\\
2.4375	0.0841946085\\
};
\addplot [color=blue, forget plot]
  table[row sep=crcr]{%
2.6875	0.0557341925\\
2.6875	0.1175657075\\
2.8125	0.1175657075\\
2.8125	0.0557341925\\
2.6875	0.0557341925\\
};
\addplot [color=blue, forget plot]
  table[row sep=crcr]{%
3.1875	0.120835236\\
3.1875	0.1243548345\\
3.3125	0.1243548345\\
3.3125	0.120835236\\
3.1875	0.120835236\\
};
\addplot [color=blue, forget plot]
  table[row sep=crcr]{%
3.4375	0.1363445055\\
3.4375	0.2574689365\\
3.5625	0.2574689365\\
3.5625	0.1363445055\\
3.4375	0.1363445055\\
};
\addplot [color=blue, forget plot]
  table[row sep=crcr]{%
3.6875	0.0858501055\\
3.6875	0.156054798\\
3.8125	0.156054798\\
3.8125	0.0858501055\\
3.6875	0.0858501055\\
};
\addplot [color=blue, forget plot]
  table[row sep=crcr]{%
4.1875	0.20060523175\\
4.1875	0.20554893375\\
4.3125	0.20554893375\\
4.3125	0.20060523175\\
4.1875	0.20060523175\\
};
\addplot [color=blue, forget plot]
  table[row sep=crcr]{%
4.4375	0.193567634\\
4.4375	0.3897081445\\
4.5625	0.3897081445\\
4.5625	0.193567634\\
4.4375	0.193567634\\
};
\addplot [color=blue, forget plot]
  table[row sep=crcr]{%
4.6875	0.1160456535\\
4.6875	0.247085482\\
4.8125	0.247085482\\
4.8125	0.1160456535\\
4.6875	0.1160456535\\
};
\addplot [color=blue, forget plot]
  table[row sep=crcr]{%
5.1875	0.25028277825\\
5.1875	0.3050969475\\
5.3125	0.3050969475\\
5.3125	0.25028277825\\
5.1875	0.25028277825\\
};
\addplot [color=blue, forget plot]
  table[row sep=crcr]{%
5.4375	0.295282608\\
5.4375	0.617178367\\
5.5625	0.617178367\\
5.5625	0.295282608\\
5.4375	0.295282608\\
};
\addplot [color=blue, forget plot]
  table[row sep=crcr]{%
5.6875	0.1890805815\\
5.6875	0.3471068555\\
5.8125	0.3471068555\\
5.8125	0.1890805815\\
5.6875	0.1890805815\\
};
\addplot [color=blue, forget plot]
  table[row sep=crcr]{%
6.1875	0.46330008025\\
6.1875	0.47848464025\\
6.3125	0.47848464025\\
6.3125	0.46330008025\\
6.1875	0.46330008025\\
};
\addplot [color=blue, forget plot]
  table[row sep=crcr]{%
6.4375	0.414276705\\
6.4375	0.87937192\\
6.5625	0.87937192\\
6.5625	0.414276705\\
6.4375	0.414276705\\
};
\addplot [color=blue, forget plot]
  table[row sep=crcr]{%
6.6875	0.2986842225\\
6.6875	0.563423231\\
6.8125	0.563423231\\
6.8125	0.2986842225\\
6.6875	0.2986842225\\
};
\addplot [color=blue, forget plot]
  table[row sep=crcr]{%
7.1875	0.651586804\\
7.1875	0.671889845\\
7.3125	0.671889845\\
7.3125	0.651586804\\
7.1875	0.651586804\\
};
\addplot [color=blue, forget plot]
  table[row sep=crcr]{%
7.4375	0.6095816415\\
7.4375	1.2144539735\\
7.5625	1.2144539735\\
7.5625	0.6095816415\\
7.4375	0.6095816415\\
};
\addplot [color=blue, forget plot]
  table[row sep=crcr]{%
7.6875	0.4037924045\\
7.6875	0.7657274065\\
7.8125	0.7657274065\\
7.8125	0.4037924045\\
7.6875	0.4037924045\\
};
\addplot [color=blue, forget plot]
  table[row sep=crcr]{%
8.1875	0.911693595\\
8.1875	0.935355293\\
8.3125	0.935355293\\
8.3125	0.911693595\\
8.1875	0.911693595\\
};
\addplot [color=blue, forget plot]
  table[row sep=crcr]{%
8.4375	0.863271304\\
8.4375	1.68311927\\
8.5625	1.68311927\\
8.5625	0.863271304\\
8.4375	0.863271304\\
};
\addplot [color=blue, forget plot]
  table[row sep=crcr]{%
8.6875	0.5489324535\\
8.6875	0.9987906975\\
8.8125	0.9987906975\\
8.8125	0.5489324535\\
8.6875	0.5489324535\\
};
\addplot [color=blue, forget plot]
  table[row sep=crcr]{%
9.1875	1.0988397155\\
9.1875	1.132957689\\
9.3125	1.132957689\\
9.3125	1.0988397155\\
9.1875	1.0988397155\\
};
\addplot [color=blue, forget plot]
  table[row sep=crcr]{%
9.4375	1.084868247\\
9.4375	2.26860556\\
9.5625	2.26860556\\
9.5625	1.084868247\\
9.4375	1.084868247\\
};
\addplot [color=blue, forget plot]
  table[row sep=crcr]{%
9.6875	0.7394289345\\
9.6875	1.3982497295\\
9.8125	1.3982497295\\
9.8125	0.7394289345\\
9.6875	0.7394289345\\
};
\addplot [color=red, forget plot]
  table[row sep=crcr]{%
1.1875	0.037354412\\
1.3125	0.037354412\\
};
\addplot [color=red, forget plot]
  table[row sep=crcr]{%
1.4375	0.068541377\\
1.5625	0.068541377\\
};
\addplot [color=red, forget plot]
  table[row sep=crcr]{%
1.6875	0.048424302\\
1.8125	0.048424302\\
};
\addplot [color=red, forget plot]
  table[row sep=crcr]{%
2.1875	0.0728312595\\
2.3125	0.0728312595\\
};
\addplot [color=red, forget plot]
  table[row sep=crcr]{%
2.4375	0.1106178175\\
2.5625	0.1106178175\\
};
\addplot [color=red, forget plot]
  table[row sep=crcr]{%
2.6875	0.0855377715\\
2.8125	0.0855377715\\
};
\addplot [color=red, forget plot]
  table[row sep=crcr]{%
3.1875	0.122448344\\
3.3125	0.122448344\\
};
\addplot [color=red, forget plot]
  table[row sep=crcr]{%
3.4375	0.169805028\\
3.5625	0.169805028\\
};
\addplot [color=red, forget plot]
  table[row sep=crcr]{%
3.6875	0.121882503\\
3.8125	0.121882503\\
};
\addplot [color=red, forget plot]
  table[row sep=crcr]{%
4.1875	0.202505999\\
4.3125	0.202505999\\
};
\addplot [color=red, forget plot]
  table[row sep=crcr]{%
4.4375	0.233480631\\
4.5625	0.233480631\\
};
\addplot [color=red, forget plot]
  table[row sep=crcr]{%
4.6875	0.182245801\\
4.8125	0.182245801\\
};
\addplot [color=red, forget plot]
  table[row sep=crcr]{%
5.1875	0.299632539\\
5.3125	0.299632539\\
};
\addplot [color=red, forget plot]
  table[row sep=crcr]{%
5.4375	0.403892082\\
5.5625	0.403892082\\
};
\addplot [color=red, forget plot]
  table[row sep=crcr]{%
5.6875	0.271028748\\
5.8125	0.271028748\\
};
\addplot [color=red, forget plot]
  table[row sep=crcr]{%
6.1875	0.468617424\\
6.3125	0.468617424\\
};
\addplot [color=red, forget plot]
  table[row sep=crcr]{%
6.4375	0.523253424\\
6.5625	0.523253424\\
};
\addplot [color=red, forget plot]
  table[row sep=crcr]{%
6.6875	0.4219599225\\
6.8125	0.4219599225\\
};
\addplot [color=red, forget plot]
  table[row sep=crcr]{%
7.1875	0.661811162\\
7.3125	0.661811162\\
};
\addplot [color=red, forget plot]
  table[row sep=crcr]{%
7.4375	0.796315642\\
7.5625	0.796315642\\
};
\addplot [color=red, forget plot]
  table[row sep=crcr]{%
7.6875	0.597279864\\
7.8125	0.597279864\\
};
\addplot [color=red, forget plot]
  table[row sep=crcr]{%
8.1875	0.919726562\\
8.3125	0.919726562\\
};
\addplot [color=red, forget plot]
  table[row sep=crcr]{%
8.4375	1.1512953585\\
8.5625	1.1512953585\\
};
\addplot [color=red, forget plot]
  table[row sep=crcr]{%
8.6875	0.7600344855\\
8.8125	0.7600344855\\
};
\addplot [color=red, forget plot]
  table[row sep=crcr]{%
9.1875	1.1137690765\\
9.3125	1.1137690765\\
};
\addplot [color=red, forget plot]
  table[row sep=crcr]{%
9.4375	1.4343450415\\
9.5625	1.4343450415\\
};
\addplot [color=red, forget plot]
  table[row sep=crcr]{%
9.6875	1.054676679\\
9.8125	1.054676679\\
};
\addplot [color=black, draw=none, mark=+, mark options={solid, red}, forget plot]
  table[row sep=crcr]{%
1.25	0.053865459\\
};
\addplot [color=black, draw=none, mark=+, mark options={solid, red}, forget plot]
  table[row sep=crcr]{%
nan	nan\\
};
\addplot [color=black, draw=none, mark=+, mark options={solid, red}, forget plot]
  table[row sep=crcr]{%
nan	nan\\
};
\addplot [color=black, draw=none, mark=+, mark options={solid, red}, forget plot]
  table[row sep=crcr]{%
2.25	0.07623932\\
2.25	0.076703754\\
2.25	0.077876551\\
};
\addplot [color=black, draw=none, mark=+, mark options={solid, red}, forget plot]
  table[row sep=crcr]{%
nan	nan\\
};
\addplot [color=black, draw=none, mark=+, mark options={solid, red}, forget plot]
  table[row sep=crcr]{%
2.75	0.218182487\\
};
\addplot [color=black, draw=none, mark=+, mark options={solid, red}, forget plot]
  table[row sep=crcr]{%
3.25	0.130752196\\
3.25	0.137473124\\
};
\addplot [color=black, draw=none, mark=+, mark options={solid, red}, forget plot]
  table[row sep=crcr]{%
nan	nan\\
};
\addplot [color=black, draw=none, mark=+, mark options={solid, red}, forget plot]
  table[row sep=crcr]{%
3.75	0.332176211\\
};
\addplot [color=black, draw=none, mark=+, mark options={solid, red}, forget plot]
  table[row sep=crcr]{%
4.25	0.214998666\\
4.25	0.215436162\\
4.25	0.216056813\\
4.25	0.216530155\\
4.25	0.217276694\\
4.25	0.217525743\\
};
\addplot [color=black, draw=none, mark=+, mark options={solid, red}, forget plot]
  table[row sep=crcr]{%
nan	nan\\
};
\addplot [color=black, draw=none, mark=+, mark options={solid, red}, forget plot]
  table[row sep=crcr]{%
4.75	0.447374947\\
4.75	0.528793249\\
};
\addplot [color=black, draw=none, mark=+, mark options={solid, red}, forget plot]
  table[row sep=crcr]{%
5.25	0.479727865\\
};
\addplot [color=black, draw=none, mark=+, mark options={solid, red}, forget plot]
  table[row sep=crcr]{%
nan	nan\\
};
\addplot [color=black, draw=none, mark=+, mark options={solid, red}, forget plot]
  table[row sep=crcr]{%
5.75	0.60671325\\
5.75	0.652638235\\
};
\addplot [color=black, draw=none, mark=+, mark options={solid, red}, forget plot]
  table[row sep=crcr]{%
6.25	0.382811137\\
6.25	0.386227681\\
6.25	0.387354866\\
6.25	0.388969432\\
6.25	0.501821231\\
6.25	0.505143793\\
6.25	0.511408539\\
6.25	0.518190873\\
};
\addplot [color=black, draw=none, mark=+, mark options={solid, red}, forget plot]
  table[row sep=crcr]{%
nan	nan\\
};
\addplot [color=black, draw=none, mark=+, mark options={solid, red}, forget plot]
  table[row sep=crcr]{%
6.75	1.047114137\\
6.75	1.079811578\\
};
\addplot [color=black, draw=none, mark=+, mark options={solid, red}, forget plot]
  table[row sep=crcr]{%
7.25	0.703080897\\
7.25	0.707208449\\
};
\addplot [color=black, draw=none, mark=+, mark options={solid, red}, forget plot]
  table[row sep=crcr]{%
nan	nan\\
};
\addplot [color=black, draw=none, mark=+, mark options={solid, red}, forget plot]
  table[row sep=crcr]{%
7.75	1.325771723\\
};
\addplot [color=black, draw=none, mark=+, mark options={solid, red}, forget plot]
  table[row sep=crcr]{%
nan	nan\\
};
\addplot [color=black, draw=none, mark=+, mark options={solid, red}, forget plot]
  table[row sep=crcr]{%
nan	nan\\
};
\addplot [color=black, draw=none, mark=+, mark options={solid, red}, forget plot]
  table[row sep=crcr]{%
8.75	2.198457841\\
};
\addplot [color=black, draw=none, mark=+, mark options={solid, red}, forget plot]
  table[row sep=crcr]{%
9.25	1.193764074\\
9.25	1.239587367\\
};
\addplot [color=black, draw=none, mark=+, mark options={solid, red}, forget plot]
  table[row sep=crcr]{%
nan	nan\\
};
\addplot [color=black, draw=none, mark=+, mark options={solid, red}, forget plot]
  table[row sep=crcr]{%
9.75	2.389652569\\
};

\addplot[area legend, draw=black, fill=red, fill opacity=0.5]
table[row sep=crcr] {%
x	y\\
9.6875	0.7394289345\\
9.6875	1.3982497295\\
9.8125	1.3982497295\\
9.8125	0.7394289345\\
9.6875	0.7394289345\\
}--cycle;
\addlegendentry{active-set}

\addplot[area legend, draw=black, fill=green, fill opacity=0.5]
table[row sep=crcr] {%
x	y\\
9.4375	1.084868247\\
9.4375	2.26860556\\
9.5625	2.26860556\\
9.5625	1.084868247\\
9.4375	1.084868247\\
}--cycle;
\addlegendentry{SQP}

\addplot[area legend, draw=black, fill=blue, fill opacity=0.5]
table[row sep=crcr] {%
x	y\\
9.1875	1.0988397155\\
9.1875	1.132957689\\
9.3125	1.132957689\\
9.3125	1.0988397155\\
9.1875	1.0988397155\\
}--cycle;
\addlegendentry{IPM}

\addplot[area legend, draw=black, fill=red, fill opacity=0.5, forget plot]
table[row sep=crcr] {%
x	y\\
8.6875	0.5489324535\\
8.6875	0.9987906975\\
8.8125	0.9987906975\\
8.8125	0.5489324535\\
8.6875	0.5489324535\\
}--cycle;

\addplot[area legend, draw=black, fill=green, fill opacity=0.5, forget plot]
table[row sep=crcr] {%
x	y\\
8.4375	0.863271304\\
8.4375	1.68311927\\
8.5625	1.68311927\\
8.5625	0.863271304\\
8.4375	0.863271304\\
}--cycle;

\addplot[area legend, draw=black, fill=blue, fill opacity=0.5, forget plot]
table[row sep=crcr] {%
x	y\\
8.1875	0.911693595\\
8.1875	0.935355293\\
8.3125	0.935355293\\
8.3125	0.911693595\\
8.1875	0.911693595\\
}--cycle;

\addplot[area legend, draw=black, fill=red, fill opacity=0.5, forget plot]
table[row sep=crcr] {%
x	y\\
7.6875	0.4037924045\\
7.6875	0.7657274065\\
7.8125	0.7657274065\\
7.8125	0.4037924045\\
7.6875	0.4037924045\\
}--cycle;

\addplot[area legend, draw=black, fill=green, fill opacity=0.5, forget plot]
table[row sep=crcr] {%
x	y\\
7.4375	0.6095816415\\
7.4375	1.2144539735\\
7.5625	1.2144539735\\
7.5625	0.6095816415\\
7.4375	0.6095816415\\
}--cycle;

\addplot[area legend, draw=black, fill=blue, fill opacity=0.5, forget plot]
table[row sep=crcr] {%
x	y\\
7.1875	0.651586804\\
7.1875	0.671889845\\
7.3125	0.671889845\\
7.3125	0.651586804\\
7.1875	0.651586804\\
}--cycle;

\addplot[area legend, draw=black, fill=red, fill opacity=0.5, forget plot]
table[row sep=crcr] {%
x	y\\
6.6875	0.2986842225\\
6.6875	0.563423231\\
6.8125	0.563423231\\
6.8125	0.2986842225\\
6.6875	0.2986842225\\
}--cycle;

\addplot[area legend, draw=black, fill=green, fill opacity=0.5, forget plot]
table[row sep=crcr] {%
x	y\\
6.4375	0.414276705\\
6.4375	0.87937192\\
6.5625	0.87937192\\
6.5625	0.414276705\\
6.4375	0.414276705\\
}--cycle;

\addplot[area legend, draw=black, fill=blue, fill opacity=0.5, forget plot]
table[row sep=crcr] {%
x	y\\
6.1875	0.46330008025\\
6.1875	0.47848464025\\
6.3125	0.47848464025\\
6.3125	0.46330008025\\
6.1875	0.46330008025\\
}--cycle;

\addplot[area legend, draw=black, fill=red, fill opacity=0.5, forget plot]
table[row sep=crcr] {%
x	y\\
5.6875	0.1890805815\\
5.6875	0.3471068555\\
5.8125	0.3471068555\\
5.8125	0.1890805815\\
5.6875	0.1890805815\\
}--cycle;

\addplot[area legend, draw=black, fill=green, fill opacity=0.5, forget plot]
table[row sep=crcr] {%
x	y\\
5.4375	0.295282608\\
5.4375	0.617178367\\
5.5625	0.617178367\\
5.5625	0.295282608\\
5.4375	0.295282608\\
}--cycle;

\addplot[area legend, draw=black, fill=blue, fill opacity=0.5, forget plot]
table[row sep=crcr] {%
x	y\\
5.1875	0.25028277825\\
5.1875	0.3050969475\\
5.3125	0.3050969475\\
5.3125	0.25028277825\\
5.1875	0.25028277825\\
}--cycle;

\addplot[area legend, draw=black, fill=red, fill opacity=0.5, forget plot]
table[row sep=crcr] {%
x	y\\
4.6875	0.1160456535\\
4.6875	0.247085482\\
4.8125	0.247085482\\
4.8125	0.1160456535\\
4.6875	0.1160456535\\
}--cycle;

\addplot[area legend, draw=black, fill=green, fill opacity=0.5, forget plot]
table[row sep=crcr] {%
x	y\\
4.4375	0.193567634\\
4.4375	0.3897081445\\
4.5625	0.3897081445\\
4.5625	0.193567634\\
4.4375	0.193567634\\
}--cycle;

\addplot[area legend, draw=black, fill=blue, fill opacity=0.5, forget plot]
table[row sep=crcr] {%
x	y\\
4.1875	0.20060523175\\
4.1875	0.20554893375\\
4.3125	0.20554893375\\
4.3125	0.20060523175\\
4.1875	0.20060523175\\
}--cycle;

\addplot[area legend, draw=black, fill=red, fill opacity=0.5, forget plot]
table[row sep=crcr] {%
x	y\\
3.6875	0.0858501055\\
3.6875	0.156054798\\
3.8125	0.156054798\\
3.8125	0.0858501055\\
3.6875	0.0858501055\\
}--cycle;

\addplot[area legend, draw=black, fill=green, fill opacity=0.5, forget plot]
table[row sep=crcr] {%
x	y\\
3.4375	0.1363445055\\
3.4375	0.2574689365\\
3.5625	0.2574689365\\
3.5625	0.1363445055\\
3.4375	0.1363445055\\
}--cycle;

\addplot[area legend, draw=black, fill=blue, fill opacity=0.5, forget plot]
table[row sep=crcr] {%
x	y\\
3.1875	0.120835236\\
3.1875	0.1243548345\\
3.3125	0.1243548345\\
3.3125	0.120835236\\
3.1875	0.120835236\\
}--cycle;

\addplot[area legend, draw=black, fill=red, fill opacity=0.5, forget plot]
table[row sep=crcr] {%
x	y\\
2.6875	0.0557341925\\
2.6875	0.1175657075\\
2.8125	0.1175657075\\
2.8125	0.0557341925\\
2.6875	0.0557341925\\
}--cycle;

\addplot[area legend, draw=black, fill=green, fill opacity=0.5, forget plot]
table[row sep=crcr] {%
x	y\\
2.4375	0.0841946085\\
2.4375	0.1752357035\\
2.5625	0.1752357035\\
2.5625	0.0841946085\\
2.4375	0.0841946085\\
}--cycle;

\addplot[area legend, draw=black, fill=blue, fill opacity=0.5, forget plot]
table[row sep=crcr] {%
x	y\\
2.1875	0.072214465\\
2.1875	0.073786595\\
2.3125	0.073786595\\
2.3125	0.072214465\\
2.1875	0.072214465\\
}--cycle;

\addplot[area legend, draw=black, fill=red, fill opacity=0.5, forget plot]
table[row sep=crcr] {%
x	y\\
1.6875	0.0341194575\\
1.6875	0.076390559\\
1.8125	0.076390559\\
1.8125	0.0341194575\\
1.6875	0.0341194575\\
}--cycle;

\addplot[area legend, draw=black, fill=green, fill opacity=0.5, forget plot]
table[row sep=crcr] {%
x	y\\
1.4375	0.050791497\\
1.4375	0.107235153\\
1.5625	0.107235153\\
1.5625	0.050791497\\
1.4375	0.050791497\\
}--cycle;

\addplot[area legend, draw=black, fill=blue, fill opacity=0.5, forget plot]
table[row sep=crcr] {%
x	y\\
1.1875	0.0367141805\\
1.1875	0.04068242\\
1.3125	0.04068242\\
1.3125	0.0367141805\\
1.1875	0.0367141805\\
}--cycle;
\end{axis}
\end{tikzpicture}%

%% file: figs/solvingtime.tex
\definecolor{mycolor1}{rgb}{0.00000,1.00000,1.00000}%
\begin{tikzpicture}

\begin{axis}[%
width=0.951\figwidth,
height=\figheight,
at={(0\figwidth,0\figheight)},
scale only axis,
unbounded coords=jump,
separate axis lines,
every outer x axis line/.append style={white!15!black},
every x tick label/.append style={font=\color{white!15!black}},
every x tick/.append style={white!15!black},
xmin=1.125,
xmax=12.125,
xtick={1.375,2.125,2.875,3.625,4.375,5.125,5.875,6.625,7.375,8.125,8.875,9.625,10.375,11.125,11.875},
xticklabels={{100},{200},{300},{400},{500},{600},{700},{800},{900},{1000},{1100},{1200},{1300},{1400},{1500}},
xlabel style={font=\color{white!15!black}},
xlabel={Training data size},
every outer y axis line/.append style={white!15!black},
every y tick label/.append style={font=\color{white!15!black}},
every y tick/.append style={white!15!black},
ymin=0,
ymax=2.9952753605,
ytick={0,0.5,1,1.5,2,2.5},
ylabel style={font=\color{white!15!black}},
ylabel={Solving time (s)},
axis background/.style={fill=white},
ymajorgrids,
grid style={solid},
legend style={at={(0.03,0.97)}, anchor=north west, legend cell align=left, align=left, draw=white!15!black}
]
\addplot [color=black, dashed, line width=0.5pt, forget plot]
  table[row sep=crcr]{%
1.25	0.088244431\\
1.25	0.093402903\\
};
\addplot [color=black, dashed, line width=0.5pt, forget plot]
  table[row sep=crcr]{%
1.5	0.036837629\\
1.5	0.038101097\\
};
\addplot [color=black, dashed, line width=0.5pt, forget plot]
  table[row sep=crcr]{%
2	0.0836333465\\
2	0.085458099\\
};
\addplot [color=black, dashed, line width=0.5pt, forget plot]
  table[row sep=crcr]{%
2.25	0.06184177425\\
2.25	0.064000632\\
};
\addplot [color=black, dashed, line width=0.5pt, forget plot]
  table[row sep=crcr]{%
2.75	0.08813033325\\
2.75	0.089703339\\
};
\addplot [color=black, dashed, line width=0.5pt, forget plot]
  table[row sep=crcr]{%
3	0.0996911345\\
3	0.103277977\\
};
\addplot [color=black, dashed, line width=0.5pt, forget plot]
  table[row sep=crcr]{%
3.5	0.092437556\\
3.5	0.09440703\\
};
\addplot [color=black, dashed, line width=0.5pt, forget plot]
  table[row sep=crcr]{%
3.75	0.1628354805\\
3.75	0.168884772\\
};
\addplot [color=black, dashed, line width=0.5pt, forget plot]
  table[row sep=crcr]{%
4.25	0.096775786\\
4.25	0.097901022\\
};
\addplot [color=black, dashed, line width=0.5pt, forget plot]
  table[row sep=crcr]{%
4.5	0.25023784\\
4.5	0.266005507\\
};
\addplot [color=black, dashed, line width=0.5pt, forget plot]
  table[row sep=crcr]{%
5	0.10392167675\\
5	0.105910282\\
};
\addplot [color=black, dashed, line width=0.5pt, forget plot]
  table[row sep=crcr]{%
5.25	0.379391802\\
5.25	0.405533701\\
};
\addplot [color=black, dashed, line width=0.5pt, forget plot]
  table[row sep=crcr]{%
5.75	0.1125318455\\
5.75	0.113991805\\
};
\addplot [color=black, dashed, line width=0.5pt, forget plot]
  table[row sep=crcr]{%
6	0.5436682495\\
6	0.565877589\\
};
\addplot [color=black, dashed, line width=0.5pt, forget plot]
  table[row sep=crcr]{%
6.5	0.123815388\\
6.5	0.125312513\\
};
\addplot [color=black, dashed, line width=0.5pt, forget plot]
  table[row sep=crcr]{%
6.75	0.758382252\\
6.75	0.786658014\\
};
\addplot [color=black, dashed, line width=0.5pt, forget plot]
  table[row sep=crcr]{%
7.25	0.134863786\\
7.25	0.13601792\\
};
\addplot [color=black, dashed, line width=0.5pt, forget plot]
  table[row sep=crcr]{%
7.5	0.9131437395\\
7.5	0.939302255\\
};
\addplot [color=black, dashed, line width=0.5pt, forget plot]
  table[row sep=crcr]{%
8	0.152843893\\
8	0.166693484\\
};
\addplot [color=black, dashed, line width=0.5pt, forget plot]
  table[row sep=crcr]{%
8.25	1.15444260925\\
8.25	1.185951459\\
};
\addplot [color=black, dashed, line width=0.5pt, forget plot]
  table[row sep=crcr]{%
8.75	0.1551369475\\
8.75	0.15807081\\
};
\addplot [color=black, dashed, line width=0.5pt, forget plot]
  table[row sep=crcr]{%
9	1.404901609\\
9	1.443482845\\
};
\addplot [color=black, dashed, line width=0.5pt, forget plot]
  table[row sep=crcr]{%
9.5	0.1660060905\\
9.5	0.16849097\\
};
\addplot [color=black, dashed, line width=0.5pt, forget plot]
  table[row sep=crcr]{%
9.75	1.76152949025\\
9.75	1.803812745\\
};
\addplot [color=black, dashed, line width=0.5pt, forget plot]
  table[row sep=crcr]{%
10.25	0.17879915875\\
10.25	0.181215358\\
};
\addplot [color=black, dashed, line width=0.5pt, forget plot]
  table[row sep=crcr]{%
10.5	2.052393913\\
10.5	2.107722001\\
};
\addplot [color=black, dashed, line width=0.5pt, forget plot]
  table[row sep=crcr]{%
11	0.19472411975\\
11	0.198478276\\
};
\addplot [color=black, dashed, line width=0.5pt, forget plot]
  table[row sep=crcr]{%
11.25	2.3968169745\\
11.25	2.46841955\\
};
\addplot [color=black, dashed, line width=0.5pt, forget plot]
  table[row sep=crcr]{%
11.75	0.209733893\\
11.75	0.212051366\\
};
\addplot [color=black, dashed, line width=0.5pt, forget plot]
  table[row sep=crcr]{%
12	2.75677745725\\
12	2.812712784\\
};
\addplot [color=black, dashed, line width=0.5pt, forget plot]
  table[row sep=crcr]{%
1.25	0.080364492\\
1.25	0.081507304\\
};
\addplot [color=black, dashed, line width=0.5pt, forget plot]
  table[row sep=crcr]{%
1.5	0.034308719\\
1.5	0.035738664\\
};
\addplot [color=black, dashed, line width=0.5pt, forget plot]
  table[row sep=crcr]{%
2	0.081301935\\
2	0.082319099\\
};
\addplot [color=black, dashed, line width=0.5pt, forget plot]
  table[row sep=crcr]{%
2.25	0.058533654\\
2.25	0.0600900835\\
};
\addplot [color=black, dashed, line width=0.5pt, forget plot]
  table[row sep=crcr]{%
2.75	0.085944419\\
2.75	0.0868941285\\
};
\addplot [color=black, dashed, line width=0.5pt, forget plot]
  table[row sep=crcr]{%
3	0.095497901\\
3	0.097114532\\
};
\addplot [color=black, dashed, line width=0.5pt, forget plot]
  table[row sep=crcr]{%
3.5	0.089552011\\
3.5	0.090983177\\
};
\addplot [color=black, dashed, line width=0.5pt, forget plot]
  table[row sep=crcr]{%
3.75	0.156339021\\
3.75	0.1584581145\\
};
\addplot [color=black, dashed, line width=0.5pt, forget plot]
  table[row sep=crcr]{%
4.25	0.094659566\\
4.25	0.095928497\\
};
\addplot [color=black, dashed, line width=0.5pt, forget plot]
  table[row sep=crcr]{%
4.5	0.235109371\\
4.5	0.239550652\\
};
\addplot [color=black, dashed, line width=0.5pt, forget plot]
  table[row sep=crcr]{%
5	0.101499555\\
5	0.10250888425\\
};
\addplot [color=black, dashed, line width=0.5pt, forget plot]
  table[row sep=crcr]{%
5.25	0.354424591\\
5.25	0.361662601\\
};
\addplot [color=black, dashed, line width=0.5pt, forget plot]
  table[row sep=crcr]{%
5.75	0.109401487\\
5.75	0.111179857\\
};
\addplot [color=black, dashed, line width=0.5pt, forget plot]
  table[row sep=crcr]{%
6	0.513897633\\
6	0.5266444695\\
};
\addplot [color=black, dashed, line width=0.5pt, forget plot]
  table[row sep=crcr]{%
6.5	0.121571171\\
6.5	0.122657595\\
};
\addplot [color=black, dashed, line width=0.5pt, forget plot]
  table[row sep=crcr]{%
6.75	0.712557005\\
6.75	0.7326691255\\
};
\addplot [color=black, dashed, line width=0.5pt, forget plot]
  table[row sep=crcr]{%
7.25	0.132254898\\
7.25	0.13358331\\
};
\addplot [color=black, dashed, line width=0.5pt, forget plot]
  table[row sep=crcr]{%
7.5	0.855590169\\
7.5	0.880989538\\
};
\addplot [color=black, dashed, line width=0.5pt, forget plot]
  table[row sep=crcr]{%
8	0.139415943\\
8	0.141756142\\
};
\addplot [color=black, dashed, line width=0.5pt, forget plot]
  table[row sep=crcr]{%
8.25	1.112409653\\
8.25	1.13225638925\\
};
\addplot [color=black, dashed, line width=0.5pt, forget plot]
  table[row sep=crcr]{%
8.75	0.150583372\\
8.75	0.153150762\\
};
\addplot [color=black, dashed, line width=0.5pt, forget plot]
  table[row sep=crcr]{%
9	1.359312283\\
9	1.377396974\\
};
\addplot [color=black, dashed, line width=0.5pt, forget plot]
  table[row sep=crcr]{%
9.5	0.162492576\\
9.5	0.1641519575\\
};
\addplot [color=black, dashed, line width=0.5pt, forget plot]
  table[row sep=crcr]{%
9.75	1.678934681\\
9.75	1.7199891235\\
};
\addplot [color=black, dashed, line width=0.5pt, forget plot]
  table[row sep=crcr]{%
10.25	0.175730649\\
10.25	0.177039733\\
};
\addplot [color=black, dashed, line width=0.5pt, forget plot]
  table[row sep=crcr]{%
10.5	1.961171026\\
10.5	2.011487981\\
};
\addplot [color=black, dashed, line width=0.5pt, forget plot]
  table[row sep=crcr]{%
11	0.190737394\\
11	0.19204686525\\
};
\addplot [color=black, dashed, line width=0.5pt, forget plot]
  table[row sep=crcr]{%
11.25	2.29920672\\
11.25	2.344823538\\
};
\addplot [color=black, dashed, line width=0.5pt, forget plot]
  table[row sep=crcr]{%
11.75	0.205276288\\
11.75	0.207079114\\
};
\addplot [color=black, dashed, line width=0.5pt, forget plot]
  table[row sep=crcr]{%
12	2.656328913\\
12	2.71120865125\\
};
\addplot [color=black, line width=0.5pt, forget plot]
  table[row sep=crcr]{%
1.21875	0.093402903\\
1.28125	0.093402903\\
};
\addplot [color=black, line width=0.5pt, forget plot]
  table[row sep=crcr]{%
1.46875	0.038101097\\
1.53125	0.038101097\\
};
\addplot [color=black, line width=0.5pt, forget plot]
  table[row sep=crcr]{%
1.96875	0.085458099\\
2.03125	0.085458099\\
};
\addplot [color=black, line width=0.5pt, forget plot]
  table[row sep=crcr]{%
2.21875	0.064000632\\
2.28125	0.064000632\\
};
\addplot [color=black, line width=0.5pt, forget plot]
  table[row sep=crcr]{%
2.71875	0.089703339\\
2.78125	0.089703339\\
};
\addplot [color=black, line width=0.5pt, forget plot]
  table[row sep=crcr]{%
2.96875	0.103277977\\
3.03125	0.103277977\\
};
\addplot [color=black, line width=0.5pt, forget plot]
  table[row sep=crcr]{%
3.46875	0.09440703\\
3.53125	0.09440703\\
};
\addplot [color=black, line width=0.5pt, forget plot]
  table[row sep=crcr]{%
3.71875	0.168884772\\
3.78125	0.168884772\\
};
\addplot [color=black, line width=0.5pt, forget plot]
  table[row sep=crcr]{%
4.21875	0.097901022\\
4.28125	0.097901022\\
};
\addplot [color=black, line width=0.5pt, forget plot]
  table[row sep=crcr]{%
4.46875	0.266005507\\
4.53125	0.266005507\\
};
\addplot [color=black, line width=0.5pt, forget plot]
  table[row sep=crcr]{%
4.96875	0.105910282\\
5.03125	0.105910282\\
};
\addplot [color=black, line width=0.5pt, forget plot]
  table[row sep=crcr]{%
5.21875	0.405533701\\
5.28125	0.405533701\\
};
\addplot [color=black, line width=0.5pt, forget plot]
  table[row sep=crcr]{%
5.71875	0.113991805\\
5.78125	0.113991805\\
};
\addplot [color=black, line width=0.5pt, forget plot]
  table[row sep=crcr]{%
5.96875	0.565877589\\
6.03125	0.565877589\\
};
\addplot [color=black, line width=0.5pt, forget plot]
  table[row sep=crcr]{%
6.46875	0.125312513\\
6.53125	0.125312513\\
};
\addplot [color=black, line width=0.5pt, forget plot]
  table[row sep=crcr]{%
6.71875	0.786658014\\
6.78125	0.786658014\\
};
\addplot [color=black, line width=0.5pt, forget plot]
  table[row sep=crcr]{%
7.21875	0.13601792\\
7.28125	0.13601792\\
};
\addplot [color=black, line width=0.5pt, forget plot]
  table[row sep=crcr]{%
7.46875	0.939302255\\
7.53125	0.939302255\\
};
\addplot [color=black, line width=0.5pt, forget plot]
  table[row sep=crcr]{%
7.96875	0.166693484\\
8.03125	0.166693484\\
};
\addplot [color=black, line width=0.5pt, forget plot]
  table[row sep=crcr]{%
8.21875	1.185951459\\
8.28125	1.185951459\\
};
\addplot [color=black, line width=0.5pt, forget plot]
  table[row sep=crcr]{%
8.71875	0.15807081\\
8.78125	0.15807081\\
};
\addplot [color=black, line width=0.5pt, forget plot]
  table[row sep=crcr]{%
8.96875	1.443482845\\
9.03125	1.443482845\\
};
\addplot [color=black, line width=0.5pt, forget plot]
  table[row sep=crcr]{%
9.46875	0.16849097\\
9.53125	0.16849097\\
};
\addplot [color=black, line width=0.5pt, forget plot]
  table[row sep=crcr]{%
9.71875	1.803812745\\
9.78125	1.803812745\\
};
\addplot [color=black, line width=0.5pt, forget plot]
  table[row sep=crcr]{%
10.21875	0.181215358\\
10.28125	0.181215358\\
};
\addplot [color=black, line width=0.5pt, forget plot]
  table[row sep=crcr]{%
10.46875	2.107722001\\
10.53125	2.107722001\\
};
\addplot [color=black, line width=0.5pt, forget plot]
  table[row sep=crcr]{%
10.96875	0.198478276\\
11.03125	0.198478276\\
};
\addplot [color=black, line width=0.5pt, forget plot]
  table[row sep=crcr]{%
11.21875	2.46841955\\
11.28125	2.46841955\\
};
\addplot [color=black, line width=0.5pt, forget plot]
  table[row sep=crcr]{%
11.71875	0.212051366\\
11.78125	0.212051366\\
};
\addplot [color=black, line width=0.5pt, forget plot]
  table[row sep=crcr]{%
11.96875	2.812712784\\
12.03125	2.812712784\\
};
\addplot [color=black, line width=0.5pt, forget plot]
  table[row sep=crcr]{%
1.21875	0.080364492\\
1.28125	0.080364492\\
};
\addplot [color=black, line width=0.5pt, forget plot]
  table[row sep=crcr]{%
1.46875	0.034308719\\
1.53125	0.034308719\\
};
\addplot [color=black, line width=0.5pt, forget plot]
  table[row sep=crcr]{%
1.96875	0.081301935\\
2.03125	0.081301935\\
};
\addplot [color=black, line width=0.5pt, forget plot]
  table[row sep=crcr]{%
2.21875	0.058533654\\
2.28125	0.058533654\\
};
\addplot [color=black, line width=0.5pt, forget plot]
  table[row sep=crcr]{%
2.71875	0.085944419\\
2.78125	0.085944419\\
};
\addplot [color=black, line width=0.5pt, forget plot]
  table[row sep=crcr]{%
2.96875	0.095497901\\
3.03125	0.095497901\\
};
\addplot [color=black, line width=0.5pt, forget plot]
  table[row sep=crcr]{%
3.46875	0.089552011\\
3.53125	0.089552011\\
};
\addplot [color=black, line width=0.5pt, forget plot]
  table[row sep=crcr]{%
3.71875	0.156339021\\
3.78125	0.156339021\\
};
\addplot [color=black, line width=0.5pt, forget plot]
  table[row sep=crcr]{%
4.21875	0.094659566\\
4.28125	0.094659566\\
};
\addplot [color=black, line width=0.5pt, forget plot]
  table[row sep=crcr]{%
4.46875	0.235109371\\
4.53125	0.235109371\\
};
\addplot [color=black, line width=0.5pt, forget plot]
  table[row sep=crcr]{%
4.96875	0.101499555\\
5.03125	0.101499555\\
};
\addplot [color=black, line width=0.5pt, forget plot]
  table[row sep=crcr]{%
5.21875	0.354424591\\
5.28125	0.354424591\\
};
\addplot [color=black, line width=0.5pt, forget plot]
  table[row sep=crcr]{%
5.71875	0.109401487\\
5.78125	0.109401487\\
};
\addplot [color=black, line width=0.5pt, forget plot]
  table[row sep=crcr]{%
5.96875	0.513897633\\
6.03125	0.513897633\\
};
\addplot [color=black, line width=0.5pt, forget plot]
  table[row sep=crcr]{%
6.46875	0.121571171\\
6.53125	0.121571171\\
};
\addplot [color=black, line width=0.5pt, forget plot]
  table[row sep=crcr]{%
6.71875	0.712557005\\
6.78125	0.712557005\\
};
\addplot [color=black, line width=0.5pt, forget plot]
  table[row sep=crcr]{%
7.21875	0.132254898\\
7.28125	0.132254898\\
};
\addplot [color=black, line width=0.5pt, forget plot]
  table[row sep=crcr]{%
7.46875	0.855590169\\
7.53125	0.855590169\\
};
\addplot [color=black, line width=0.5pt, forget plot]
  table[row sep=crcr]{%
7.96875	0.139415943\\
8.03125	0.139415943\\
};
\addplot [color=black, line width=0.5pt, forget plot]
  table[row sep=crcr]{%
8.21875	1.112409653\\
8.28125	1.112409653\\
};
\addplot [color=black, line width=0.5pt, forget plot]
  table[row sep=crcr]{%
8.71875	0.150583372\\
8.78125	0.150583372\\
};
\addplot [color=black, line width=0.5pt, forget plot]
  table[row sep=crcr]{%
8.96875	1.359312283\\
9.03125	1.359312283\\
};
\addplot [color=black, line width=0.5pt, forget plot]
  table[row sep=crcr]{%
9.46875	0.162492576\\
9.53125	0.162492576\\
};
\addplot [color=black, line width=0.5pt, forget plot]
  table[row sep=crcr]{%
9.71875	1.678934681\\
9.78125	1.678934681\\
};
\addplot [color=black, line width=0.5pt, forget plot]
  table[row sep=crcr]{%
10.21875	0.175730649\\
10.28125	0.175730649\\
};
\addplot [color=black, line width=0.5pt, forget plot]
  table[row sep=crcr]{%
10.46875	1.961171026\\
10.53125	1.961171026\\
};
\addplot [color=black, line width=0.5pt, forget plot]
  table[row sep=crcr]{%
10.96875	0.190737394\\
11.03125	0.190737394\\
};
\addplot [color=black, line width=0.5pt, forget plot]
  table[row sep=crcr]{%
11.21875	2.29920672\\
11.28125	2.29920672\\
};
\addplot [color=black, line width=0.5pt, forget plot]
  table[row sep=crcr]{%
11.71875	0.205276288\\
11.78125	0.205276288\\
};
\addplot [color=black, line width=0.5pt, forget plot]
  table[row sep=crcr]{%
11.96875	2.656328913\\
12.03125	2.656328913\\
};
\addplot [color=blue, line width=0.5pt, forget plot]
  table[row sep=crcr]{%
1.1875	0.081507304\\
1.1875	0.088244431\\
1.3125	0.088244431\\
1.3125	0.081507304\\
1.1875	0.081507304\\
};
\addplot [color=blue, line width=0.5pt, forget plot]
  table[row sep=crcr]{%
1.4375	0.035738664\\
1.4375	0.036837629\\
1.5625	0.036837629\\
1.5625	0.035738664\\
1.4375	0.035738664\\
};
\addplot [color=blue, line width=0.5pt, forget plot]
  table[row sep=crcr]{%
1.9375	0.082319099\\
1.9375	0.0836333465\\
2.0625	0.0836333465\\
2.0625	0.082319099\\
1.9375	0.082319099\\
};
\addplot [color=blue, line width=0.5pt, forget plot]
  table[row sep=crcr]{%
2.1875	0.0600900835\\
2.1875	0.06184177425\\
2.3125	0.06184177425\\
2.3125	0.0600900835\\
2.1875	0.0600900835\\
};
\addplot [color=blue, line width=0.5pt, forget plot]
  table[row sep=crcr]{%
2.6875	0.0868941285\\
2.6875	0.08813033325\\
2.8125	0.08813033325\\
2.8125	0.0868941285\\
2.6875	0.0868941285\\
};
\addplot [color=blue, line width=0.5pt, forget plot]
  table[row sep=crcr]{%
2.9375	0.097114532\\
2.9375	0.0996911345\\
3.0625	0.0996911345\\
3.0625	0.097114532\\
2.9375	0.097114532\\
};
\addplot [color=blue, line width=0.5pt, forget plot]
  table[row sep=crcr]{%
3.4375	0.090983177\\
3.4375	0.092437556\\
3.5625	0.092437556\\
3.5625	0.090983177\\
3.4375	0.090983177\\
};
\addplot [color=blue, line width=0.5pt, forget plot]
  table[row sep=crcr]{%
3.6875	0.1584581145\\
3.6875	0.1628354805\\
3.8125	0.1628354805\\
3.8125	0.1584581145\\
3.6875	0.1584581145\\
};
\addplot [color=blue, line width=0.5pt, forget plot]
  table[row sep=crcr]{%
4.1875	0.095928497\\
4.1875	0.096775786\\
4.3125	0.096775786\\
4.3125	0.095928497\\
4.1875	0.095928497\\
};
\addplot [color=blue, line width=0.5pt, forget plot]
  table[row sep=crcr]{%
4.4375	0.239550652\\
4.4375	0.25023784\\
4.5625	0.25023784\\
4.5625	0.239550652\\
4.4375	0.239550652\\
};
\addplot [color=blue, line width=0.5pt, forget plot]
  table[row sep=crcr]{%
4.9375	0.10250888425\\
4.9375	0.10392167675\\
5.0625	0.10392167675\\
5.0625	0.10250888425\\
4.9375	0.10250888425\\
};
\addplot [color=blue, line width=0.5pt, forget plot]
  table[row sep=crcr]{%
5.1875	0.361662601\\
5.1875	0.379391802\\
5.3125	0.379391802\\
5.3125	0.361662601\\
5.1875	0.361662601\\
};
\addplot [color=blue, line width=0.5pt, forget plot]
  table[row sep=crcr]{%
5.6875	0.111179857\\
5.6875	0.1125318455\\
5.8125	0.1125318455\\
5.8125	0.111179857\\
5.6875	0.111179857\\
};
\addplot [color=blue, line width=0.5pt, forget plot]
  table[row sep=crcr]{%
5.9375	0.5266444695\\
5.9375	0.5436682495\\
6.0625	0.5436682495\\
6.0625	0.5266444695\\
5.9375	0.5266444695\\
};
\addplot [color=blue, line width=0.5pt, forget plot]
  table[row sep=crcr]{%
6.4375	0.122657595\\
6.4375	0.123815388\\
6.5625	0.123815388\\
6.5625	0.122657595\\
6.4375	0.122657595\\
};
\addplot [color=blue, line width=0.5pt, forget plot]
  table[row sep=crcr]{%
6.6875	0.7326691255\\
6.6875	0.758382252\\
6.8125	0.758382252\\
6.8125	0.7326691255\\
6.6875	0.7326691255\\
};
\addplot [color=blue, line width=0.5pt, forget plot]
  table[row sep=crcr]{%
7.1875	0.13358331\\
7.1875	0.134863786\\
7.3125	0.134863786\\
7.3125	0.13358331\\
7.1875	0.13358331\\
};
\addplot [color=blue, line width=0.5pt, forget plot]
  table[row sep=crcr]{%
7.4375	0.880989538\\
7.4375	0.9131437395\\
7.5625	0.9131437395\\
7.5625	0.880989538\\
7.4375	0.880989538\\
};
\addplot [color=blue, line width=0.5pt, forget plot]
  table[row sep=crcr]{%
7.9375	0.141756142\\
7.9375	0.152843893\\
8.0625	0.152843893\\
8.0625	0.141756142\\
7.9375	0.141756142\\
};
\addplot [color=blue, line width=0.5pt, forget plot]
  table[row sep=crcr]{%
8.1875	1.13225638925\\
8.1875	1.15444260925\\
8.3125	1.15444260925\\
8.3125	1.13225638925\\
8.1875	1.13225638925\\
};
\addplot [color=blue, line width=0.5pt, forget plot]
  table[row sep=crcr]{%
8.6875	0.153150762\\
8.6875	0.1551369475\\
8.8125	0.1551369475\\
8.8125	0.153150762\\
8.6875	0.153150762\\
};
\addplot [color=blue, line width=0.5pt, forget plot]
  table[row sep=crcr]{%
8.9375	1.377396974\\
8.9375	1.404901609\\
9.0625	1.404901609\\
9.0625	1.377396974\\
8.9375	1.377396974\\
};
\addplot [color=blue, line width=0.5pt, forget plot]
  table[row sep=crcr]{%
9.4375	0.1641519575\\
9.4375	0.1660060905\\
9.5625	0.1660060905\\
9.5625	0.1641519575\\
9.4375	0.1641519575\\
};
\addplot [color=blue, line width=0.5pt, forget plot]
  table[row sep=crcr]{%
9.6875	1.7199891235\\
9.6875	1.76152949025\\
9.8125	1.76152949025\\
9.8125	1.7199891235\\
9.6875	1.7199891235\\
};
\addplot [color=blue, line width=0.5pt, forget plot]
  table[row sep=crcr]{%
10.1875	0.177039733\\
10.1875	0.17879915875\\
10.3125	0.17879915875\\
10.3125	0.177039733\\
10.1875	0.177039733\\
};
\addplot [color=blue, line width=0.5pt, forget plot]
  table[row sep=crcr]{%
10.4375	2.011487981\\
10.4375	2.052393913\\
10.5625	2.052393913\\
10.5625	2.011487981\\
10.4375	2.011487981\\
};
\addplot [color=blue, line width=0.5pt, forget plot]
  table[row sep=crcr]{%
10.9375	0.19204686525\\
10.9375	0.19472411975\\
11.0625	0.19472411975\\
11.0625	0.19204686525\\
10.9375	0.19204686525\\
};
\addplot [color=blue, line width=0.5pt, forget plot]
  table[row sep=crcr]{%
11.1875	2.344823538\\
11.1875	2.3968169745\\
11.3125	2.3968169745\\
11.3125	2.344823538\\
11.1875	2.344823538\\
};
\addplot [color=blue, line width=0.5pt, forget plot]
  table[row sep=crcr]{%
11.6875	0.207079114\\
11.6875	0.209733893\\
11.8125	0.209733893\\
11.8125	0.207079114\\
11.6875	0.207079114\\
};
\addplot [color=blue, line width=0.5pt, forget plot]
  table[row sep=crcr]{%
11.9375	2.71120865125\\
11.9375	2.75677745725\\
12.0625	2.75677745725\\
12.0625	2.71120865125\\
11.9375	2.71120865125\\
};
\addplot [color=red, line width=0.5pt, forget plot]
  table[row sep=crcr]{%
1.1875	0.0828007895\\
1.3125	0.0828007895\\
};
\addplot [color=red, line width=0.5pt, forget plot]
  table[row sep=crcr]{%
1.4375	0.036394575\\
1.5625	0.036394575\\
};
\addplot [color=red, line width=0.5pt, forget plot]
  table[row sep=crcr]{%
1.9375	0.082885628\\
2.0625	0.082885628\\
};
\addplot [color=red, line width=0.5pt, forget plot]
  table[row sep=crcr]{%
2.1875	0.06080248\\
2.3125	0.06080248\\
};
\addplot [color=red, line width=0.5pt, forget plot]
  table[row sep=crcr]{%
2.6875	0.087342181\\
2.8125	0.087342181\\
};
\addplot [color=red, line width=0.5pt, forget plot]
  table[row sep=crcr]{%
2.9375	0.0979711095\\
3.0625	0.0979711095\\
};
\addplot [color=red, line width=0.5pt, forget plot]
  table[row sep=crcr]{%
3.4375	0.091353982\\
3.5625	0.091353982\\
};
\addplot [color=red, line width=0.5pt, forget plot]
  table[row sep=crcr]{%
3.6875	0.1604204745\\
3.8125	0.1604204745\\
};
\addplot [color=red, line width=0.5pt, forget plot]
  table[row sep=crcr]{%
4.1875	0.0962596665\\
4.3125	0.0962596665\\
};
\addplot [color=red, line width=0.5pt, forget plot]
  table[row sep=crcr]{%
4.4375	0.242944827\\
4.5625	0.242944827\\
};
\addplot [color=red, line width=0.5pt, forget plot]
  table[row sep=crcr]{%
4.9375	0.102967653\\
5.0625	0.102967653\\
};
\addplot [color=red, line width=0.5pt, forget plot]
  table[row sep=crcr]{%
5.1875	0.3681629605\\
5.3125	0.3681629605\\
};
\addplot [color=red, line width=0.5pt, forget plot]
  table[row sep=crcr]{%
5.6875	0.111754753\\
5.8125	0.111754753\\
};
\addplot [color=red, line width=0.5pt, forget plot]
  table[row sep=crcr]{%
5.9375	0.535862026\\
6.0625	0.535862026\\
};
\addplot [color=red, line width=0.5pt, forget plot]
  table[row sep=crcr]{%
6.4375	0.1232872995\\
6.5625	0.1232872995\\
};
\addplot [color=red, line width=0.5pt, forget plot]
  table[row sep=crcr]{%
6.6875	0.746146002\\
6.8125	0.746146002\\
};
\addplot [color=red, line width=0.5pt, forget plot]
  table[row sep=crcr]{%
7.1875	0.1342560695\\
7.3125	0.1342560695\\
};
\addplot [color=red, line width=0.5pt, forget plot]
  table[row sep=crcr]{%
7.4375	0.8926918645\\
7.5625	0.8926918645\\
};
\addplot [color=red, line width=0.5pt, forget plot]
  table[row sep=crcr]{%
7.9375	0.143280239\\
8.0625	0.143280239\\
};
\addplot [color=red, line width=0.5pt, forget plot]
  table[row sep=crcr]{%
8.1875	1.145006785\\
8.3125	1.145006785\\
};
\addplot [color=red, line width=0.5pt, forget plot]
  table[row sep=crcr]{%
8.6875	0.1539475115\\
8.8125	0.1539475115\\
};
\addplot [color=red, line width=0.5pt, forget plot]
  table[row sep=crcr]{%
8.9375	1.392214882\\
9.0625	1.392214882\\
};
\addplot [color=red, line width=0.5pt, forget plot]
  table[row sep=crcr]{%
9.4375	0.1650975355\\
9.5625	0.1650975355\\
};
\addplot [color=red, line width=0.5pt, forget plot]
  table[row sep=crcr]{%
9.6875	1.741321203\\
9.8125	1.741321203\\
};
\addplot [color=red, line width=0.5pt, forget plot]
  table[row sep=crcr]{%
10.1875	0.177791122\\
10.3125	0.177791122\\
};
\addplot [color=red, line width=0.5pt, forget plot]
  table[row sep=crcr]{%
10.4375	2.0356606035\\
10.5625	2.0356606035\\
};
\addplot [color=red, line width=0.5pt, forget plot]
  table[row sep=crcr]{%
10.9375	0.193124171\\
11.0625	0.193124171\\
};
\addplot [color=red, line width=0.5pt, forget plot]
  table[row sep=crcr]{%
11.1875	2.373831793\\
11.3125	2.373831793\\
};
\addplot [color=red, line width=0.5pt, forget plot]
  table[row sep=crcr]{%
11.6875	0.2081855615\\
11.8125	0.2081855615\\
};
\addplot [color=red, line width=0.5pt, forget plot]
  table[row sep=crcr]{%
11.9375	2.734118116\\
12.0625	2.734118116\\
};
\addplot [color=black, line width=0.5pt, draw=none, mark size=3.0pt, mark=+, mark options={solid, red}, forget plot]
  table[row sep=crcr]{%
1.25	0.05161213\\
1.25	0.051910337\\
1.25	0.053018302\\
1.25	0.062241509\\
1.25	0.100549961\\
1.25	0.109955877\\
1.25	0.110456284\\
1.25	0.110629888\\
1.25	0.110736233\\
1.25	0.111358146\\
1.25	0.11145422\\
1.25	0.11149077\\
1.25	0.111978065\\
1.25	0.112113463\\
1.25	0.112246606\\
1.25	0.112384817\\
1.25	0.112596576\\
1.25	0.116594361\\
1.25	0.118050351\\
1.25	0.118201012\\
1.25	0.121601421\\
1.25	0.125022195\\
};
\addplot [color=black, line width=0.5pt, draw=none, mark size=3.0pt, mark=+, mark options={solid, red}, forget plot]
  table[row sep=crcr]{%
1.5	0.040332286\\
1.5	0.040570059\\
1.5	0.040656451\\
1.5	0.04093912\\
1.5	0.041558487\\
1.5	0.0416523\\
};
\addplot [color=black, line width=0.5pt, draw=none, mark size=3.0pt, mark=+, mark options={solid, red}, forget plot]
  table[row sep=crcr]{%
2	0.086659231\\
2	0.086768187\\
2	0.087319312\\
2	0.087635541\\
2	0.095353766\\
};
\addplot [color=black, line width=0.5pt, draw=none, mark size=3.0pt, mark=+, mark options={solid, red}, forget plot]
  table[row sep=crcr]{%
2.25	0.065382203\\
2.25	0.065458411\\
2.25	0.065481258\\
2.25	0.065966232\\
2.25	0.067045083\\
2.25	0.068527301\\
};
\addplot [color=black, line width=0.5pt, draw=none, mark size=3.0pt, mark=+, mark options={solid, red}, forget plot]
  table[row sep=crcr]{%
2.75	0.090442945\\
2.75	0.091246422\\
2.75	0.091306441\\
2.75	0.091480135\\
2.75	0.092460228\\
2.75	0.092804967\\
2.75	0.09345594\\
2.75	0.094786556\\
2.75	0.09544304\\
};
\addplot [color=black, line width=0.5pt, draw=none, mark size=3.0pt, mark=+, mark options={solid, red}, forget plot]
  table[row sep=crcr]{%
3	0.103827623\\
3	0.104081987\\
3	0.104289157\\
3	0.104378466\\
3	0.106354748\\
3	0.106607173\\
3	0.106741564\\
3	0.108605405\\
3	0.108704088\\
};
\addplot [color=black, line width=0.5pt, draw=none, mark size=3.0pt, mark=+, mark options={solid, red}, forget plot]
  table[row sep=crcr]{%
3.5	0.095138825\\
3.5	0.095574781\\
3.5	0.096544475\\
3.5	0.100937957\\
};
\addplot [color=black, line width=0.5pt, draw=none, mark size=3.0pt, mark=+, mark options={solid, red}, forget plot]
  table[row sep=crcr]{%
3.75	0.169788175\\
3.75	0.170737446\\
3.75	0.173126293\\
3.75	0.176180796\\
};
\addplot [color=black, line width=0.5pt, draw=none, mark size=3.0pt, mark=+, mark options={solid, red}, forget plot]
  table[row sep=crcr]{%
4.25	0.094649801\\
4.25	0.09841491\\
4.25	0.098422414\\
4.25	0.100149664\\
4.25	0.100971954\\
4.25	0.101054536\\
};
\addplot [color=black, line width=0.5pt, draw=none, mark size=3.0pt, mark=+, mark options={solid, red}, forget plot]
  table[row sep=crcr]{%
4.5	0.266972525\\
};
\addplot [color=black, line width=0.5pt, draw=none, mark size=3.0pt, mark=+, mark options={solid, red}, forget plot]
  table[row sep=crcr]{%
5	0.106057481\\
5	0.106923501\\
5	0.107822472\\
5	0.108473154\\
5	0.108516499\\
5	0.108716799\\
5	0.111711982\\
};
\addplot [color=black, line width=0.5pt, draw=none, mark size=3.0pt, mark=+, mark options={solid, red}, forget plot]
  table[row sep=crcr]{%
5.25	0.406042975\\
5.25	0.406652291\\
5.25	0.419623119\\
5.25	0.444520927\\
5.25	0.483409044\\
};
\addplot [color=black, line width=0.5pt, draw=none, mark size=3.0pt, mark=+, mark options={solid, red}, forget plot]
  table[row sep=crcr]{%
5.75	0.114588934\\
5.75	0.114755752\\
5.75	0.114968059\\
5.75	0.11544448\\
5.75	0.117856812\\
5.75	0.118668866\\
};
\addplot [color=black, line width=0.5pt, draw=none, mark size=3.0pt, mark=+, mark options={solid, red}, forget plot]
  table[row sep=crcr]{%
6	0.573731864\\
6	0.576438863\\
6	0.583171644\\
6	0.589819786\\
6	0.61262463\\
};
\addplot [color=black, line width=0.5pt, draw=none, mark size=3.0pt, mark=+, mark options={solid, red}, forget plot]
  table[row sep=crcr]{%
6.5	0.125859815\\
6.5	0.12611764\\
6.5	0.126520442\\
6.5	0.12672866\\
6.5	0.128336914\\
6.5	0.13102434\\
6.5	0.132190405\\
6.5	0.132514906\\
6.5	0.139880965\\
6.5	0.139949309\\
};
\addplot [color=black, line width=0.5pt, draw=none, mark size=3.0pt, mark=+, mark options={solid, red}, forget plot]
  table[row sep=crcr]{%
6.75	0.807055179\\
};
\addplot [color=black, line width=0.5pt, draw=none, mark size=3.0pt, mark=+, mark options={solid, red}, forget plot]
  table[row sep=crcr]{%
7.25	0.13713438\\
7.25	0.137817337\\
7.25	0.138088873\\
7.25	0.138564628\\
7.25	0.138581578\\
};
\addplot [color=black, line width=0.5pt, draw=none, mark size=3.0pt, mark=+, mark options={solid, red}, forget plot]
  table[row sep=crcr]{%
nan	nan\\
};
\addplot [color=black, line width=0.5pt, draw=none, mark size=3.0pt, mark=+, mark options={solid, red}, forget plot]
  table[row sep=crcr]{%
8	0.091968494\\
8	0.092202962\\
8	0.092756711\\
8	0.095547155\\
8	0.098451771\\
8	0.189476325\\
8	0.189777571\\
8	0.190979244\\
8	0.190986841\\
8	0.191237473\\
8	0.19185815\\
8	0.192209459\\
8	0.192233661\\
8	0.192372598\\
8	0.193344125\\
8	0.193589598\\
8	0.193631479\\
8	0.193856367\\
8	0.194533024\\
8	0.197693301\\
8	0.198114448\\
8	0.198736111\\
8	0.204067129\\
8	0.204200116\\
8	0.207343067\\
8	0.207795443\\
};
\addplot [color=black, line width=0.5pt, draw=none, mark size=3.0pt, mark=+, mark options={solid, red}, forget plot]
  table[row sep=crcr]{%
8.25	1.098301399\\
8.25	1.203347615\\
8.25	1.206210928\\
8.25	1.23658864\\
8.25	1.244213974\\
};
\addplot [color=black, line width=0.5pt, draw=none, mark size=3.0pt, mark=+, mark options={solid, red}, forget plot]
  table[row sep=crcr]{%
8.75	0.158152013\\
8.75	0.158965335\\
8.75	0.159320323\\
8.75	0.160282402\\
8.75	0.160764707\\
8.75	0.162647988\\
};
\addplot [color=black, line width=0.5pt, draw=none, mark size=3.0pt, mark=+, mark options={solid, red}, forget plot]
  table[row sep=crcr]{%
nan	nan\\
};
\addplot [color=black, line width=0.5pt, draw=none, mark size=3.0pt, mark=+, mark options={solid, red}, forget plot]
  table[row sep=crcr]{%
9.5	0.171304547\\
9.5	0.173399064\\
9.5	0.174174131\\
9.5	0.177238687\\
9.5	0.17942779\\
9.5	0.180886416\\
9.5	0.182974587\\
9.5	0.183787489\\
};
\addplot [color=black, line width=0.5pt, draw=none, mark size=3.0pt, mark=+, mark options={solid, red}, forget plot]
  table[row sep=crcr]{%
nan	nan\\
};
\addplot [color=black, line width=0.5pt, draw=none, mark size=3.0pt, mark=+, mark options={solid, red}, forget plot]
  table[row sep=crcr]{%
10.25	0.184251403\\
10.25	0.185019747\\
10.25	0.187044327\\
10.25	0.190389579\\
10.25	0.193957322\\
};
\addplot [color=black, line width=0.5pt, draw=none, mark size=3.0pt, mark=+, mark options={solid, red}, forget plot]
  table[row sep=crcr]{%
10.5	2.113810506\\
10.5	2.114398637\\
10.5	2.205466796\\
};
\addplot [color=black, line width=0.5pt, draw=none, mark size=3.0pt, mark=+, mark options={solid, red}, forget plot]
  table[row sep=crcr]{%
11	0.208489558\\
};
\addplot [color=black, line width=0.5pt, draw=none, mark size=3.0pt, mark=+, mark options={solid, red}, forget plot]
  table[row sep=crcr]{%
11.25	2.48214018\\
11.25	2.485661517\\
};
\addplot [color=black, line width=0.5pt, draw=none, mark size=3.0pt, mark=+, mark options={solid, red}, forget plot]
  table[row sep=crcr]{%
11.75	0.213873018\\
11.75	0.214331775\\
11.75	0.217503089\\
11.75	0.222520074\\
11.75	0.223718188\\
11.75	0.223785876\\
11.75	0.223897071\\
11.75	0.224659644\\
};
\addplot [color=black, line width=0.5pt, draw=none, mark size=3.0pt, mark=+, mark options={solid, red}, forget plot]
  table[row sep=crcr]{%
12	2.840701841\\
12	2.850827734\\
12	2.854276949\\
};

\addplot[area legend, line width=0.5pt, draw=black, fill=mycolor1, fill opacity=0.5]
table[row sep=crcr] {%
x	y\\
11.6875	0.207079114\\
11.6875	0.209733893\\
11.8125	0.209733893\\
11.8125	0.207079114\\
11.6875	0.207079114\\
}--cycle;
\addlegendentry{linGP-SCP}

\addplot[area legend, line width=0.5pt, draw=black, fill=red, fill opacity=0.5]
table[row sep=crcr] {%
x	y\\
11.9375	2.71120865125\\
11.9375	2.75677745725\\
12.0625	2.75677745725\\
12.0625	2.71120865125\\
11.9375	2.71120865125\\
}--cycle;
\addlegendentry{Knitro}

\addplot[area legend, line width=0.5pt, draw=black, fill=red, fill opacity=0.5, forget plot]
table[row sep=crcr] {%
x	y\\
11.1875	2.344823538\\
11.1875	2.3968169745\\
11.3125	2.3968169745\\
11.3125	2.344823538\\
11.1875	2.344823538\\
}--cycle;

\addplot[area legend, line width=0.5pt, draw=black, fill=mycolor1, fill opacity=0.5, forget plot]
table[row sep=crcr] {%
x	y\\
10.9375	0.19204686525\\
10.9375	0.19472411975\\
11.0625	0.19472411975\\
11.0625	0.19204686525\\
10.9375	0.19204686525\\
}--cycle;

\addplot[area legend, line width=0.5pt, draw=black, fill=red, fill opacity=0.5, forget plot]
table[row sep=crcr] {%
x	y\\
10.4375	2.011487981\\
10.4375	2.052393913\\
10.5625	2.052393913\\
10.5625	2.011487981\\
10.4375	2.011487981\\
}--cycle;

\addplot[area legend, line width=0.5pt, draw=black, fill=mycolor1, fill opacity=0.5, forget plot]
table[row sep=crcr] {%
x	y\\
10.1875	0.177039733\\
10.1875	0.17879915875\\
10.3125	0.17879915875\\
10.3125	0.177039733\\
10.1875	0.177039733\\
}--cycle;

\addplot[area legend, line width=0.5pt, draw=black, fill=red, fill opacity=0.5, forget plot]
table[row sep=crcr] {%
x	y\\
9.6875	1.7199891235\\
9.6875	1.76152949025\\
9.8125	1.76152949025\\
9.8125	1.7199891235\\
9.6875	1.7199891235\\
}--cycle;

\addplot[area legend, line width=0.5pt, draw=black, fill=mycolor1, fill opacity=0.5, forget plot]
table[row sep=crcr] {%
x	y\\
9.4375	0.1641519575\\
9.4375	0.1660060905\\
9.5625	0.1660060905\\
9.5625	0.1641519575\\
9.4375	0.1641519575\\
}--cycle;

\addplot[area legend, line width=0.5pt, draw=black, fill=red, fill opacity=0.5, forget plot]
table[row sep=crcr] {%
x	y\\
8.9375	1.377396974\\
8.9375	1.404901609\\
9.0625	1.404901609\\
9.0625	1.377396974\\
8.9375	1.377396974\\
}--cycle;

\addplot[area legend, line width=0.5pt, draw=black, fill=mycolor1, fill opacity=0.5, forget plot]
table[row sep=crcr] {%
x	y\\
8.6875	0.153150762\\
8.6875	0.1551369475\\
8.8125	0.1551369475\\
8.8125	0.153150762\\
8.6875	0.153150762\\
}--cycle;

\addplot[area legend, line width=0.5pt, draw=black, fill=red, fill opacity=0.5, forget plot]
table[row sep=crcr] {%
x	y\\
8.1875	1.13225638925\\
8.1875	1.15444260925\\
8.3125	1.15444260925\\
8.3125	1.13225638925\\
8.1875	1.13225638925\\
}--cycle;

\addplot[area legend, line width=0.5pt, draw=black, fill=mycolor1, fill opacity=0.5, forget plot]
table[row sep=crcr] {%
x	y\\
7.9375	0.141756142\\
7.9375	0.152843893\\
8.0625	0.152843893\\
8.0625	0.141756142\\
7.9375	0.141756142\\
}--cycle;

\addplot[area legend, line width=0.5pt, draw=black, fill=red, fill opacity=0.5, forget plot]
table[row sep=crcr] {%
x	y\\
7.4375	0.880989538\\
7.4375	0.9131437395\\
7.5625	0.9131437395\\
7.5625	0.880989538\\
7.4375	0.880989538\\
}--cycle;

\addplot[area legend, line width=0.5pt, draw=black, fill=mycolor1, fill opacity=0.5, forget plot]
table[row sep=crcr] {%
x	y\\
7.1875	0.13358331\\
7.1875	0.134863786\\
7.3125	0.134863786\\
7.3125	0.13358331\\
7.1875	0.13358331\\
}--cycle;

\addplot[area legend, line width=0.5pt, draw=black, fill=red, fill opacity=0.5, forget plot]
table[row sep=crcr] {%
x	y\\
6.6875	0.7326691255\\
6.6875	0.758382252\\
6.8125	0.758382252\\
6.8125	0.7326691255\\
6.6875	0.7326691255\\
}--cycle;

\addplot[area legend, line width=0.5pt, draw=black, fill=mycolor1, fill opacity=0.5, forget plot]
table[row sep=crcr] {%
x	y\\
6.4375	0.122657595\\
6.4375	0.123815388\\
6.5625	0.123815388\\
6.5625	0.122657595\\
6.4375	0.122657595\\
}--cycle;

\addplot[area legend, line width=0.5pt, draw=black, fill=red, fill opacity=0.5, forget plot]
table[row sep=crcr] {%
x	y\\
5.9375	0.5266444695\\
5.9375	0.5436682495\\
6.0625	0.5436682495\\
6.0625	0.5266444695\\
5.9375	0.5266444695\\
}--cycle;

\addplot[area legend, line width=0.5pt, draw=black, fill=mycolor1, fill opacity=0.5, forget plot]
table[row sep=crcr] {%
x	y\\
5.6875	0.111179857\\
5.6875	0.1125318455\\
5.8125	0.1125318455\\
5.8125	0.111179857\\
5.6875	0.111179857\\
}--cycle;

\addplot[area legend, line width=0.5pt, draw=black, fill=red, fill opacity=0.5, forget plot]
table[row sep=crcr] {%
x	y\\
5.1875	0.361662601\\
5.1875	0.379391802\\
5.3125	0.379391802\\
5.3125	0.361662601\\
5.1875	0.361662601\\
}--cycle;

\addplot[area legend, line width=0.5pt, draw=black, fill=mycolor1, fill opacity=0.5, forget plot]
table[row sep=crcr] {%
x	y\\
4.9375	0.10250888425\\
4.9375	0.10392167675\\
5.0625	0.10392167675\\
5.0625	0.10250888425\\
4.9375	0.10250888425\\
}--cycle;

\addplot[area legend, line width=0.5pt, draw=black, fill=red, fill opacity=0.5, forget plot]
table[row sep=crcr] {%
x	y\\
4.4375	0.239550652\\
4.4375	0.25023784\\
4.5625	0.25023784\\
4.5625	0.239550652\\
4.4375	0.239550652\\
}--cycle;

\addplot[area legend, line width=0.5pt, draw=black, fill=mycolor1, fill opacity=0.5, forget plot]
table[row sep=crcr] {%
x	y\\
4.1875	0.095928497\\
4.1875	0.096775786\\
4.3125	0.096775786\\
4.3125	0.095928497\\
4.1875	0.095928497\\
}--cycle;

\addplot[area legend, line width=0.5pt, draw=black, fill=red, fill opacity=0.5, forget plot]
table[row sep=crcr] {%
x	y\\
3.6875	0.1584581145\\
3.6875	0.1628354805\\
3.8125	0.1628354805\\
3.8125	0.1584581145\\
3.6875	0.1584581145\\
}--cycle;

\addplot[area legend, line width=0.5pt, draw=black, fill=mycolor1, fill opacity=0.5, forget plot]
table[row sep=crcr] {%
x	y\\
3.4375	0.090983177\\
3.4375	0.092437556\\
3.5625	0.092437556\\
3.5625	0.090983177\\
3.4375	0.090983177\\
}--cycle;

\addplot[area legend, line width=0.5pt, draw=black, fill=red, fill opacity=0.5, forget plot]
table[row sep=crcr] {%
x	y\\
2.9375	0.097114532\\
2.9375	0.0996911345\\
3.0625	0.0996911345\\
3.0625	0.097114532\\
2.9375	0.097114532\\
}--cycle;

\addplot[area legend, line width=0.5pt, draw=black, fill=mycolor1, fill opacity=0.5, forget plot]
table[row sep=crcr] {%
x	y\\
2.6875	0.0868941285\\
2.6875	0.08813033325\\
2.8125	0.08813033325\\
2.8125	0.0868941285\\
2.6875	0.0868941285\\
}--cycle;

\addplot[area legend, line width=0.5pt, draw=black, fill=red, fill opacity=0.5, forget plot]
table[row sep=crcr] {%
x	y\\
2.1875	0.0600900835\\
2.1875	0.06184177425\\
2.3125	0.06184177425\\
2.3125	0.0600900835\\
2.1875	0.0600900835\\
}--cycle;

\addplot[area legend, line width=0.5pt, draw=black, fill=mycolor1, fill opacity=0.5, forget plot]
table[row sep=crcr] {%
x	y\\
1.9375	0.082319099\\
1.9375	0.0836333465\\
2.0625	0.0836333465\\
2.0625	0.082319099\\
1.9375	0.082319099\\
}--cycle;

\addplot[area legend, line width=0.5pt, draw=black, fill=red, fill opacity=0.5, forget plot]
table[row sep=crcr] {%
x	y\\
1.4375	0.035738664\\
1.4375	0.036837629\\
1.5625	0.036837629\\
1.5625	0.035738664\\
1.4375	0.035738664\\
}--cycle;

\addplot[area legend, line width=0.5pt, draw=black, fill=mycolor1, fill opacity=0.5, forget plot]
table[row sep=crcr] {%
x	y\\
1.1875	0.081507304\\
1.1875	0.088244431\\
1.3125	0.088244431\\
1.3125	0.081507304\\
1.1875	0.081507304\\
}--cycle;
\end{axis}
\end{tikzpicture}%

%% file: figs/solvingtime_median.tex
\begin{tikzpicture}

\begin{axis}[%
width=0.951\figwidth,
height=\figheight,
at={(0\figwidth,0\figheight)},
scale only axis,
xmin=0,
xmax=1500,
xlabel style={font=\color{white!15!black}},
xlabel={Training data size ($N$)},
ymin=0,
ymax=4.5,
ylabel style={font=\color{white!15!black}},
ylabel={Median %
  time (s)},
axis background/.style={fill=white},
xmajorgrids,
ymajorgrids,
legend style={at={(0.03,0.97)}, anchor=north west, legend cell align=left, align=left, draw=white!15!black}
]
\addplot [color=red, line width=1.2pt]
  table[row sep=crcr]{%
100	0.0327355455\\
200	0.070951205\\
300	0.128614749\\
400	0.219994723\\
500	0.3465683\\
600	0.546452678\\
700	0.78806168\\
800	1.100067258\\
900	1.348412107\\
1000	1.719321848\\
1100	2.100213798\\
1200	2.649412791\\
1300	3.090109765\\
1400	3.644734598\\
1500	4.220721804\\
};
\addlegendentry{Ipopt}

\addplot [color=green, line width=1.2pt]
  table[row sep=crcr]{%
100	0.036394575\\
200	0.06080248\\
300	0.0979711095\\
400	0.1604204745\\
500	0.242944827\\
600	0.3681629605\\
700	0.535862026\\
800	0.746146002\\
900	0.8926918645\\
1000	1.145006785\\
1100	1.392214882\\
1200	1.741321203\\
1300	2.0356606035\\
1400	2.373831793\\
1500	2.734118116\\
};
\addlegendentry{Knitro}

\addplot [color=blue, line width=1.2pt]
  table[row sep=crcr]{%
100	0.0828007895\\
200	0.082885628\\
300	0.087342181\\
400	0.091353982\\
500	0.0962596665\\
600	0.102967653\\
700	0.111754753\\
800	0.1232872995\\
900	0.1342560695\\
1000	0.143280239\\
1100	0.1539475115\\
1200	0.1650975355\\
1300	0.177791122\\
1400	0.193124171\\
1500	0.2081855615\\
};
\addlegendentry{linGP-SCP}

\end{axis}
\end{tikzpicture}%

%% file: figs/solvingtime_median_ratio.tex
\begin{tikzpicture}

\begin{axis}[%
width=0.951\figwidth,
height=\figheight,
at={(0\figwidth,0\figheight)},
scale only axis,
xmin=0,
xmax=1500,
xlabel style={font=\color{white!15!black}},
xlabel={Training data size ($N$)},
ymin=0,
ymax=25,
ylabel style={font=\color{white!15!black}},
ylabel={Rel.\ time %
  vs. linGP%
},
axis background/.style={fill=white},
xmajorgrids,
ymajorgrids,
legend style={at={(0.03,0.97)}, anchor=north west, legend cell align=left, align=left, draw=white!15!black}
]
\addplot [color=red, line width=1.2pt]
  table[row sep=crcr]{%
100	0.395353060009168\\
200	0.856013360965305\\
300	1.47253878398113\\
400	2.40815690989803\\
500	3.60034802322944\\
600	5.30703247164427\\
700	7.05170615875282\\
800	8.92279466304637\\
900	10.0435839662355\\
1000	11.9997137079036\\
1100	13.6424017350875\\
1200	16.0475611157745\\
1300	17.3805628213539\\
1400	18.8724931691746\\
1500	20.2738449947692\\
};
\addlegendentry{Ipopt}

\addplot [color=green, line width=1.2pt]
  table[row sep=crcr]{%
100	0.439543816185472\\
200	0.733570842945173\\
300	1.12169295955639\\
400	1.75603154879445\\
500	2.52384862563387\\
600	3.57552056178264\\
700	4.79498197271305\\
800	6.05209137539751\\
900	6.64917323905419\\
1000	7.99137964168248\\
1100	9.04343869176476\\
1200	10.5472271147258\\
1300	11.4497314635317\\
1400	12.2917384225302\\
1500	13.1330823151249\\
};
\addlegendentry{Knitro}

\end{axis}
\end{tikzpicture}%

%% file: conclusion.tex
We developed a fast and scalable approach to solve a general class of GP-based MPC problems.
The approach is based on the concept of \emph{linearized Gaussian Process} (\linGP), proposed in this paper, and Sequential Convex Programming.
Our approach not only solves GP-MPC faster than other NLP methods but also is much less influenced by the GP training data size -- a key factor affecting the computational complexity of GPs and GP-MPC.
Therefore, our algorithm is more scalable and predictable, which makes it more suitable for real-time control.

We are developing a more efficient implementation of the method %
and will conduct real-time control experiments.
We are also studying the convergence properties of the linGP-SCP algorithm, and investigating different applications of the \linGP concept other than MPC.

%% file: main.bbl
\begin{thebibliography}{10}
\providecommand{\url}[1]{#1}
\csname url@rmstyle\endcsname
\providecommand{\newblock}{\relax}
\providecommand{\bibinfo}[2]{#2}
\providecommand\BIBentrySTDinterwordspacing{\spaceskip=0pt\relax}
\providecommand\BIBentryALTinterwordstretchfactor{4}
\providecommand\BIBentryALTinterwordspacing{\spaceskip=\fontdimen2\font plus
\BIBentryALTinterwordstretchfactor\fontdimen3\font minus
  \fontdimen4\font\relax}
\providecommand\BIBforeignlanguage[2]{{%
\expandafter\ifx\csname l@#1\endcsname\relax
\typeout{** WARNING: IEEEtran.bst: No hyphenation pattern has been}%
\typeout{** loaded for the language `#1'. Using the pattern for}%
\typeout{** the default language instead.}%
\else
\language=\csname l@#1\endcsname
\fi
#2}}

\bibitem{maciejowski_predictive_2002}
J.~M. Maciejowski, \emph{Predictive control: with constraints}.\hskip 1em plus
  0.5em minus 0.4em\relax Pearson, 2002.

\bibitem{schwarm1999Chanceconstrainedmodelpredictive}
A.~T. Schwarm and M.~Nikolaou, ``Chance-constrained model predictive control,''
  \emph{AIChE Journal}, vol.~45, no.~8, pp. 1743--1752, Aug. 1999.

\bibitem{mayne2014Modelpredictivecontrol}
D.~Q. Mayne, ``Model predictive control: {{Recent}} developments and future
  promise,'' \emph{Automatica}, vol.~50, no.~12, pp. 2967--2986, 2014.

\bibitem{JainICCPS2018}
A.~Jain, T.~X. Nghiem, M.~Morari, and R.~Mangharam, ``{L}earning and {C}ontrol
  using {G}aussian {P}rocesses: Towards bridging machine learning and controls
  for physical systems,'' in \emph{Proc. of the International Conference on
  Cyber-Physical Systems (ICCPS)}.\hskip 1em plus 0.5em minus 0.4em\relax
  ACM/IEEE, 2018.

\bibitem{Kocijan2016}
J.~Kocijan, \emph{Modelling and control of dynamic systems using Gaussian
  process models}.\hskip 1em plus 0.5em minus 0.4em\relax Springer, 2016.

\bibitem{Rasmussen2006}
C.~E. Rasmussen and C.~K. Williams, \emph{Gaussian processes for machine
  learning}.\hskip 1em plus 0.5em minus 0.4em\relax MIT press Cambridge, 2006,
  vol.~1.

\bibitem{nghiemetal16gp}
T.~X. Nghiem and C.~N. Jones, ``Data-driven demand response modeling and
  control of buildings with gaussian processes,'' in \emph{Proceedings of
  American Control Conference (ACC)}, 2017.

\bibitem{cao2017Gaussianprocessmodel}
G.~Cao, E.~M.-K. Lai, and F.~Alam, ``Gaussian process model predictive control
  of unknown non-linear systems,'' \emph{IET Control Theory \& Applications},
  vol.~11, no.~5, pp. 703--713, Mar. 2017.

\bibitem{mao2016Successiveconvexificationnonconvex}
Y.~Mao, M.~Szmuk, and B.~A{\c c}\i{}kme{\c s}e, ``Successive convexification of
  non-convex optimal control problems and its convergence properties,'' in
  \emph{{IEEE} Conference on Decision and Control ({CDC})}, 2016.

\bibitem{lawrynczukComputationallyEfficientModel2014}
M.~Lawrynczuk, \emph{Computationally {{Efficient Model Predictive Control
  Algorithms}}: {{A Neural Network Approach}}}, ser. Studies in {{Systems}},
  {{Decision}} and {{Control}}.\hskip 1em plus 0.5em minus 0.4em\relax
  {Springer International Publishing}, 2014.

\bibitem{solak2003DerivativeObservationsGaussian}
E.~Solak, R.~{Murray-smith}, W.~E. Leithead, D.~J. Leith, and C.~E. Rasmussen,
  ``Derivative {{Observations}} in {{Gaussian Process Models}} of {{Dynamic
  Systems}},'' in \emph{Advances in {{Neural Information Processing Systems}}
  15}.\hskip 1em plus 0.5em minus 0.4em\relax {MIT Press}, 2003, pp.
  1057--1064.

\bibitem{Andersson2018}
J.~A.~E. Andersson, J.~Gillis, G.~Horn, J.~B. Rawlings, and M.~Diehl,
  ``{CasADi} -- {A} software framework for nonlinear optimization and optimal
  control,'' \emph{Mathematical Programming Computation}, 2018.

\bibitem{Waechter2009b}
A.~W{\"a}chter and L.~Biegler, ``Ipopt-an interior point optimizer,'' 2009.

\bibitem{Yalmip2004}
J.~L{\"{o}}fberg, ``Yalmip : A toolbox for modeling and optimization in
  matlab,'' in \emph{In Proc. of the CACSD Conference}, Taipei, Taiwan, 2004.

\end{thebibliography}
